\documentclass[11pt,english,authoryear]{article}
\usepackage[T1]{fontenc}
\usepackage[latin9]{inputenc}
\usepackage{geometry}
\usepackage{array}
\usepackage{verbatim}
\usepackage{float}
\usepackage{booktabs}
\usepackage{bm}
\usepackage{amsmath}
\usepackage{amsthm}
\usepackage{amssymb}
\usepackage{graphicx}
\usepackage{setspace}
\usepackage[authoryear]{natbib}
\usepackage{color}
\makeatletter

\providecommand{\tabularnewline}{\\}

\newcommand{\e}{\mathbb{E}}
\newcommand{\der}{\,\mbox{d}}
\newcommand*{\defeq}{\mathrel{\vcenter{\baselineskip0.5ex \lineskiplimit0pt \hbox{\scriptsize.}\hbox{\scriptsize.}}} =}
\usepackage{bm, bbm}

\makeatother

\usepackage{babel}
\usepackage{authblk}

\begin{document}

\title{Evaluating the Discrimination Ability of Proper Multivariate Scoring Rules}

\author{C. Alexander$\dag^*$, M. Coulon$\dag $, Y. Han${\S}$, X. Meng$\dag$\affil{$\dag$Department of Accounting and Finance, University of Sussex Business School, Falmer, Brighton BN1 9SL, United Kingdom}
}

\maketitle

\begin{abstract}
	\noindent Proper scoring rules are commonly applied to quantify the accuracy
	of distribution forecasts. Given an observation  they assign a scalar score to each distribution forecast, with the 
	the lowest expected score attributed to the true distribution. The energy and variogram scores are two
rules that have recently gained some popularity in multivariate
	settings because 
	their computation does not require a forecast to have
	parametric density function and so they are broadly applicable. Here we conduct a simulation study to compare the discrimination ability between the energy score and three variogram scores.
	Compared with other studies, our simulation
	design is more realistic because it is supported by a
	historical data set containing commodity prices, currencies and interest rates, and our data generating processes 
include a diverse selection
	of models with different marginal distributions, dependence structure, 
	and calibration windows. This facilitates a comprehensive
	comparison of the performance of proper scoring rules in different settings. To compare the scores we use three metrics: the mean relative score, error rate and a generalised discrimination heuristic. Overall, we find that the variogram score with parameter $p=0.5$ outperforms the energy score and the other two variogram scores.
\end{abstract}


\footnotesize
\noindent $\dag$ Department of Accounting and Finance, University of Sussex Business School, Falmer, Brighton BN1 9SL, UK\\
${\S}$ McKinsey and Co., Frankfurt, Germany\\
$^*$ Peking University PHBS Business School
\newpage

\normalsize
\section{Introduction}\label{sec:Introduction}

Alongside the profusion of models for generating point or distribution
forecasts, a prolific strand of theoretical research focusses
on developing methods for evaluating these forecasts. %
A standard approach is to quantify the accuracy of each prediction
with a proper scoring rule (see, for example, \citealt{gneiting2007b}).
Many scoring rules are proposed and considered in the literature:
\citet{bao2007} advocate using the Kullback--Leibler information
criterion which is derived from the logarithmic score; \citet{gneiting2007b}
advocate using the continuous ranked probability score (CRPS) and
\citet{gneiting2011} extend this to adopt the weighting approach
of \citet{amisano2007} so that evaluation can be focused on a specific
area of the distribution, such as a tail or the centre; \citet{gneiting2007b}
generalise the CRPS to the energy score for multivariate distributions;
\citet{scheuerer2015} use the concept of variograms from geostatistics
to derive a proper score; \citet{hyvarinen2005estimation} and \citet{parry2012proper} consider local scoring rules, which are based on the density function or its derivatives; \citet{diks2011likelihood} and \citet{diks2014} generalise the log score to conditional likelihood and censored likelihood, which can allocate more weights on specific regions.

Given the large number of scoring rules, conventional wisdom dictates
to apply a suitable one for the application at hand. While it is generally
agreed upon that only proper scoring rules quantify the accuracy of
probabilistic forecasts adequately \citep{winkler1996,gneiting2011},
the question of which of the proper scores to use remains largely
open \citep{gneiting2007b}. This problem is especially relevant for
multivariate evaluation, since the rankings of univariate scoring
rules mostly coincide, which reduces the risk of conflicting conclusions
(see, for example, \citealp{vonHolstein1970,winkler1971,bickel2007}).

Some previous studies provide an analytic assessment of proper scoring rules  for a specific
forecasting problem, but these are restricted to a univariate setting
in which some strong assumptions are made (see, for example, \citealt{machete2013,buja2005,merkle2013,johnstone2011}).
Previous studies comparing multivariate proper scoring rules are limited to very simple simulation settings. In their evaluation of the energy score, \citet{pinson2013}  restrict themselves to a bivariate Gaussian as the true distribution. They introduce a useful discrimination heuristic
	which measures the average relative distance of the sub-optimal scores (i.e. those  obtained from mis-specified models) 
	from the score obtained from the true distribution. They conclude that the energy score
is able to discriminate errors in mean but lacks sensitivity
to errors in variance and especially to errors in correlation.\footnote{Even in
the worst case considered, where a perfect correlation is mistaken
for zero-correlation, the energy score changes only by 7\%.} \citet{scheuerer2015} compare three different variogram scores with the energy score and the score proposed by \cite{dawid1999}. Their system has 5 or 15 dimensions but again they only consider a very simple true model, i.e. a Gaussian or
 Poisson  distribution. A simulation study assesses
the ability of each score to identify the correct model through a simple heuristic, i.e. the
score sample mean. Overall, only the variogram score
with $p=0.5$ is devoid of ranking issues in their study. Also, they confirm
the finding of \citet{pinson2013} that the energy score lacks sensitivity
to misspecification in the dependency structure.  \citet{ziel2019multivariate} compare the sensitivity of the energy
	score, variogram score and the score proposed by \citet{dawid1999}.
	They first consider a simulation study focusing on multivariate Gaussian distributions of unit variances and different correlations. Then they consider a time series empirical study where autoregressive
	(AR) models with different lags are applied to airline passengers.
	They conclude that the energy score performs the best in their simulation
	study, which seems to contradict the conclusion from \citet{pinson2013}.
	\citet{Diks2020531} use conditional likelihood and censored likelihood
	to compare multivariate and univariate approaches for forecasting
	portfolio returns. Their empirical study considers the univariate
	and multivariate generalised autoregressive conditional heteroscedasticity (GARCH) models with elliptical distributional assumptions
	applied to financial data. Their focus is not to compare scoring rules,
	but to illustrate that multivariate models do not necessarily lead
	to better forecasting accuracy for portfolio returns when compared
	with a univariate approach.

We make two major contributions to this literature. The first is to propose new metrics for discriminating between correctly specified and misspecified models using proper multivariate scoring rules. The second is to provide the first empirical study on real financial data of the relative performance of multivariate scoring rules. Most empirical research in the forecasting literature in finance and economics
includes a short empirical study, but extensive application of proper
scoring rules, even to univariate distribution forecasts of financial
returns, are hard to find. We fill this gap in the literature 
by conducting an extensive comparison between different 
rules across a diverse selection of eight static and dynamic models for multivariate distributions. We do this in a forecasting setting using three large sets of daily financial data spanning 1991 to 2018: commodity prices, exchange rates and interest rates.  Each data set has eight dimensions. We apply our new metrics to draw conclusions about the ability of  different proper scoring rules  to detect model misspecification. We do this by calibrating each model to one of the datasets and then simulating day-ahead values using the calibrated model. The exercise is repeated quarterly across many years and, every time, results are generated with each of eight different models as the true data generation process. An online appendix contains further results which have been omitted here for lack of space.

In the following: Section \ref{sec:Rules} describes the scoring rules; Section \ref{sec:Models} describes the models for multivariate distribution predictions; Section \ref{sec:Simulation} describes the design of the experiment; Section \ref{sec:Results} presents our results and Section \ref{sec:Conclusions} concludes. 

\section{Review of Scoring Rules}\label{sec:Rules}

Forecasting accuracy evaluations rely on loss measures to quantify
the performance of a distribution forecast. As \citet{diebold1995}
mention, this loss generally depends on the underlying economic structures
associated with the forecast. Scoring rules offer a promising measure
by condensing the accuracy of a distribution forecast to a single
penalty oriented value while retaining attractive statistical properties.

Let $\mathcal{F}$ be the convex class of distributions on $(\Omega,\mathcal{A})$.
Let $y$ be an observation of the random variable $Y$ and let $F$
be a forecast of the distribution of $Y$. A scoring rule is a function
\begin{align*}
S(F,y):\mathcal{F}\times\Omega\longrightarrow\mathbb{R}\cup\{-\infty,\infty\}
\end{align*}
A scoring rule $S$ is said to be proper if for all distributions $F$ and the true distribution
$G$, the following holds, 
\begin{equation}
\e_{G}S(G,\bullet)\leq\e_{G}S(F,\bullet).\label{Equation:DefinitionProper}
\end{equation}
Further, a scoring rule is strictly proper if equation~(\ref{Equation:DefinitionProper})
holds if $G$ is the unique minimiser.

Propriety of a scoring rule is important because the ideal forecast
is preferred irrespective of the cost-loss structure \citep{diebold1998,granger2000}.
A proper scoring rule is designed so that a forecaster who believes
the future distribution to be $G$ has no incentive to predict any
distribution $F\neq G$ \citep{gneiting2007a}. The term has been
coined by \citet{winkler1996,winkler1977} who shows that proper scoring
rules test for both calibration and sharpness of a distribution forecast
simultaneously. The usage of non-proper scoring rules is generally
not recommended since those can lead to wrong inferences \citep{gneiting2011}. 

The most well-known score is undoubtedly the log score, defined as
follows,
\[
\mbox{LogS}(F,y)=-\log(f(y)),
\]
where $f$ is the probability density function (PDF) of the distribution $F$. The log score has
been widely applied in both model evaluation and estimation. In the
log score, the PDF $f$ clearly cannot have value 0. To overcome
this weakness, quadratic and pseudo-spherical scores have been considered
as alternative PDF-based scores
\begin{eqnarray*}
\mbox{QS}(F,y) & = & \Vert f\Vert_{2}^{2}-2f(y),\\
\mbox{PseudoS}(F,y) & = & f(y)^{\alpha-1}/\Vert f\Vert_{\alpha}^{\alpha-1}.
\end{eqnarray*}
where $\Vert f\Vert_{\alpha}\defeq\left(\int f(y)^{\alpha}\mu\der y\right)^{1/\alpha}.$
The spherical score is a special case of the pseudo-spherical score
with $\alpha=2$. All three scores are strictly proper. However, these
scores have been criticised for only crediting forecasts for high
probabilities of the realizing value but not for high probabilities
of values near the realizing one \citep{gneiting2007b}. In addition,
the PDF-based scores rely on predictive densities which might not
be available, especially with ensemble forecasts. In view of these
limitations, efforts have been devoted to developing alternative scores
that do not require predictive densities. In the rest of section,
we briefly review three such scores: the CRPS, energy score and variogram
score.  

\subsection{Conditional Ranked Probability Score}

The CRPS introduced by \citet{matheson1976} and augmented by \citet{gneiting2011}
has been widely used to properly compare distribution forecasts with
a potential focus on certain regions of interest. The CRPS is defined
as 
\begin{align}
\mbox{CRPS}_{\nu}(F,y) & =\int_{0}^{1}\mbox{QS}_{\alpha}(F^{-1}(\alpha),y)\nu(\alpha)\der\alpha,\label{eq:CRPS_quantile}\\
\mbox{QS}_{\alpha}(F^{-1}(\alpha),y) & =2(\mathbbm{1}\{y\leq F^{-1}(\alpha)\}-\alpha)(F^{-1}(\alpha)-y)\nonumber 
\end{align}
where $F^{-1}(\alpha)=\inf\{x|F(x)=\alpha\}$ denotes the $\alpha$ quantile of $F$, 
 $\nu:[0,1]\rightarrow\mathbb{R}_{\geq0}$
is a quantile weight function and $\mbox{QS}_{\alpha}$ is known as
the quantile score.

\begin{figure}[!htb]
\caption{CRPS Schematic}
\label{Figure:CRPSSchematic} \centering \includegraphics[width=1\textwidth]{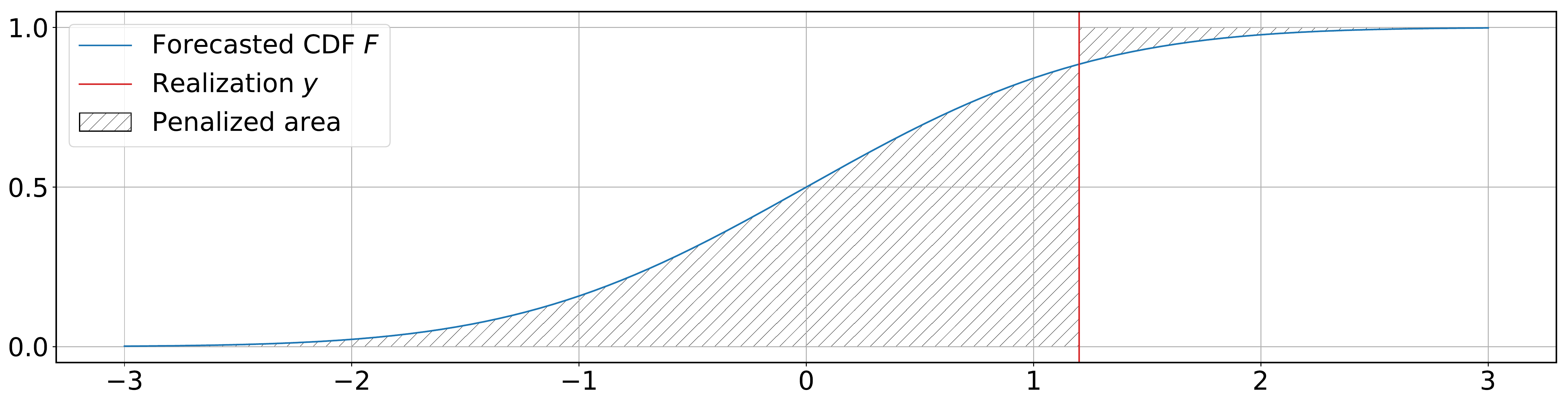} 
\raggedright{}{\footnotesize{}We use $F=\Phi$ the standard normal distribution function, and $y=1.2$ to illustrate
the concept of the CRPS. The forecasted distribution $F$ is penalized
for the shaded area left and right of the realized value $y$ through
${\int_{-\infty}^{y}F(z)^{2}\der z}$ and ${\int_{y}^{\infty}(1-F(z))^{2}\der z}$
respectively. A low score suggests high sharpness of the distribution
forecast around the realisation. 
.}{\footnotesize\par}
\end{figure}

Apart from this quantile score representation, CRPS can also be expressed
using the Brier probability score through 
\begin{align}
\mbox{CRPS}_{u}(F,y) & =\int_{-\infty}^{\infty}\mbox{PS}(F(z),\mathbbm{1}\{y\leq z\})u(z)\der z,\label{Equation:CRPSRepresentation}\\
\mbox{PS}(F(z),\mathbbm{1}\{y\leq z\}) & =\left(F(z)-\mathbbm{1}\{y\leq z\}\right)^{2}\nonumber 
\end{align}
with threshold weight function $u:\mathbb{R}\rightarrow\mathbb{R}_{\geq0}$
as shown by \citet{laio2007}. Given a realization $y$, the integral
of equation~(\ref{Equation:CRPSRepresentation}) splits into two
easily interpretable parts which get penalized by the score as visualized
in Figure~\ref{Figure:CRPSSchematic}. We remark that equations~(\ref{eq:CRPS_quantile})
and (\ref{Equation:CRPSRepresentation}) are generally not equivalent.

Additionally, \citet{gneiting2007b} derive the kernel score representation
\begin{align}
\mbox{CRPS}_{u}(F,y) & =\e_{F}\left(Y-y\right)-\frac{1}{2}\e_{F}\left(Y-Y'\right),\label{Equation:CRPSKernelRepresentation}
\end{align}
where $Y$ and $Y'$ are independent random variables with sampling
distribution $F$. This concise expression serves as a foundation
for the generalization of CRPS to the multivariate energy score discussed
in the next subsection.

For distributions with finite first moment, the CRPS is strictly proper.
Thus, the true probability function receives the lowest CRPS and is
preferred to any other probabilistic forecast. Emphasizing specific
parts of the distribution by the choice of the quantile or threshold
weight functions is simple since any non-negative functions $\nu$
and $u$ can be used, provided that equations~(\ref{eq:CRPS_quantile})
and (\ref{Equation:CRPSRepresentation}) are convergent. Table~\ref{Table:CRPSWeights}
lists the proposed functions by \citet{amisano2007} that we use in our analysis.  

\begin{table}[!htb]
\centering {\small{}\caption{Possible weight functions s for CRPS}
\label{Table:CRPSWeights} }%
\begin{tabular}{p{2.5cm}p{4cm}p{4.5cm}}
\toprule 
{\small{}Emphasis} & {\small{}Quantile weights} & {\small{}Threshold weights}\tabularnewline
\midrule 
{\small{}Uniform} & {\small{}$\nu(\alpha)=1$} & {\small{}$u(z)=1$}\tabularnewline
{\small{}Centre} & {\small{}$\nu(\alpha)=\alpha(1-\alpha)$} & {\small{}$u(z)=\varphi(z)$}\tabularnewline
{\small{}Both tails} & {\small{}$\nu(\alpha)=(2\alpha-1)^{2}$} & {\small{}$u(z)=1-\varphi(z)/\varphi(0)$}\tabularnewline
{\small{}Right tail} & {\small{}$\nu(\alpha)=\alpha^{2}$} & {\small{}$u(z)=\Phi(z)$}\tabularnewline
{\small{}Left tail} & {\small{}$\nu(\alpha)=(1-\alpha)^{2}$} & {\small{}$u(z)=1-\Phi(z)$}\tabularnewline
\bottomrule
\end{tabular}{\small{}\medskip{}
}{\small\par}
\raggedright{}{\footnotesize{}The weight functions $\nu:[0,1]\rightarrow\mathbb{R}_{\geq0}$
and $u:\mathbb{R}\rightarrow\mathbb{R}_{\geq0}$ put additional emphasis
on certain parts of the distribution. Forecasts which deviate on those
parts are penalized additionally and receive a higher CRPS. Here,
$\varphi$ and $\Phi$ denote the density and the distribution functions of
the standard normal distribution.}{\footnotesize\par}
\end{table}

\subsection{Energy Score}

The energy score is a popular multivariate strictly proper score introduced
by \citet{gneiting2007b} which generalizes the kernel representation
of the CRPS in equation~(\ref{Equation:CRPSKernelRepresentation}). Let
$\mathbf{y}=(y_{1},\ldots,y_{d})'$ be an observation of the random
vector $\mathbf{Y}$ and let $F$ be a forecast of the distribution
of $\mathbf{Y}$ such that $\e_{F}(\Vert\mathbf{Y}\Vert^{\beta})$
is finite. The energy score is then defined as 
\[
\mbox{ES}_{\beta}(F,\mathbf{y})=\frac{1}{2}\e_{F}\left(\Vert\mathbf{Y}-\mathbf{Y}'\Vert^{\beta}\right)-\e_{F}\left(\Vert\mathbf{Y-\mathbf{y}}\Vert^{\beta}\right),
\]
where $\mathbf{Y}$ and $\mathbf{Y}'$ are independent random vectors
with distribution $F$.  

\citet{szekely2003} shows that the energy score with $\beta\in(0,2)$
is strictly proper while \citet{gneiting2007b} provide an alternative
general proof. In practice, $\beta=1$ has been a common choice. In
this case, the energy score reduces to the CRPS in the univariate
case. Closed form expressions of the energy score are generally unavailable
which means that computations are done through Monte Carlo methods.

Despite its popularity, this score has been criticized for being insensitive
to misspecification of the dependency structure \citep{pinson2012,pinson2013}
and for being unable to distinguish a good representation of the predictive
distribution from a very sparse one \citep{scheuerer2015}. 

\subsection{Variogram Score}

The variogram score proposed by \citet{scheuerer2015} is based on
the concept of variograms from geostatistics. Similar to diagnostic
methods by \citet{hamill2001} and \citet{feldmann2015}, the score
examines pairwise element differences of the variables $y_{i}$ of
$\mathbf{y}$. The variogram score of order $p$ is defined as
\begin{align*}
\mbox{VS}_{p}(F,\mathbf{y})=\sum_{i=1}^{d}\sum_{j=1}^{d}\left(|y_{i}-y_{j}|^{p}-\e_{F}\left(|Y_{i}-Y_{j}|^{p}\right)\right)^{2},
\end{align*}
where $Y_{i}$ and $Y_{j}$ are the $i$-th and $j$-th components
of a random vector $\mathbf{Y}$ with distribution $F$. Intuitively,
the score makes use of the variogram of order $p$ ,
\begin{align*}
\gamma_{p}(i,j)=\frac{1}{2}\e\left(|Y_{i}-Y_{j}|^{p}\right),
\end{align*}
which quantifies the degree of spatial dependence of a stochastic
process. Pairwise comparisons measure the closeness of the deviations
in the observations relative to those of the corresponding expectations.

\citet{scheuerer2015} show that the score is proper relative to the
class of distributions for which the $2p$-th moments of all elements
are finite. The variogram score is not strictly proper because it
depends only on the $p$-th absolute moment of the distribution of
the element differences. Therefore, it cannot distinguish any distributions
where the element differences deviate in higher moments but are the same for moments of order less than or equal to 
$p$.

\begin{figure}[!htb]
\caption{Variogram observation of various orders}
\label{Figure:Variogram} \centering \includegraphics[width=1\textwidth]{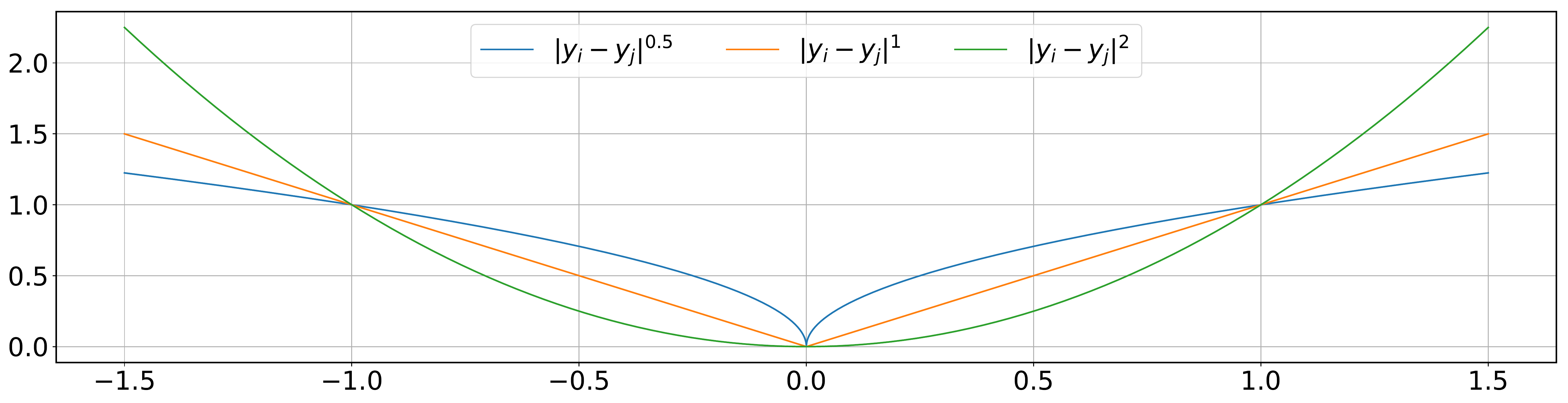} 
\raggedright{}{\footnotesize{}The figure shows the effect of the variogram
order depending on the observed absolute difference $|y_{i}-y_{j}|$.
Slight deviations in $|y_{i}-y_{j}|$ affect the observed variogram
$|y_{i}-y_{j}|^{p}$ differently, depending on its order $p$.}{\footnotesize\par}
\end{figure}

In practice, the choice of $p$ depends on the forecasted distribution
and should generally be large enough to consider all relevant moments
of the pairwise deviations but not too large to overly emphasize outliers
through the exponentiation. The values $p=0.5,1,2$ are often suggested, and Figure~\ref{Figure:Variogram} shows
the effect of $p$ by illustrating the observed variogram $|y_{i}-y_{j}|^{p}$
of different popular orders relative to changes in $|y_{i}-y_{j}|$.
It is clearly visible that the magnitude of the effect depends heavily
on the value of $|y_{i}-y_{j}|$ with the absolute slope varying between
$0$ and $3$ in the depicted domain $(-1.5,1.5)$. The sensitivity
of the observed variogram in a neighbourhood of zero deviation is
strongest for $p=0.5$ and very weak for $p=2$. This order reverses
for $|y_{i}-y_{j}|>(1/4)^{2/3}$.  We expect parameter
$p=0.5$ to be more sensitive to similar $y_{i}$ and $y_{j}$
while $p=2$ will respond more to cases where $|y_{i}-y_{j}|$ is expected
to be large. The choice $p=1$ should be able to strike a balance between these two cases. 

As with the energy score, the encapsulation of the information into
a single score leads to a loss of information. However, empirical
applications support the sensitivity of the score to flawed forecasts,
especially regarding the dependency structure \citep{scheuerer2015}.
Since the score is based on pairwise deviations, any bias that is
the same for all components of the forecast cancels out and is therefore
undetectable. This further motivates the good practice to using multiple
proper scores for the evaluation of multivariate distribution forecasts.

\section{Models for Predicting Multivariate Distributions}\label{sec:Models}

While most prior studies of this type have only considered static parametric distributions with fixed parameters
	as data generating process (DGP) and parameter variations for the misspecified models, we aim here to instead select from a broad range of realistic models for both the DGP and the misspecified models.  There is a very long list of alternative static and dynamic models that could be selected for our study.  We require a diverse selection but would like all of these to fit the historical datasets reasonably well, while differing in features such as whether the marginal distributions are static, whether the dependence structure is static, how much of the noise is filtered from the historical data,\footnote{Here noise refers to the variation in the data which is not useful for improving forecast error, similarly to the defintion of noise that can be identified by Principal Component Analysis (PCA).} and finally how much sampling error is generated by the length of the calibration window.  In this section we summarize all the models to be used in our study, beginning with static models before moving to dynamic models. 
	
	\subsection{Static Models}\label{sec:FQ}
Our simplest model assumes that the next period joint distribution of the variables can be well approximated by their joint historical distribution, i.e. the frequency distribution of contemporaneous observations on the random variables. The lack of complexity makes it this approach particularly popular with practitioners. In a survey by \cite{perignon2010} of  risk  models used by commercial banks, almost 3/4 used such historical simulation techniques. Moreover,  \cite{Danielsson2016} shows that it remains unclear whether dynamic time-series models can outperform this so-called `historical' approach, at least for predicting quantiles. We therefore include models that use historical data to build an empirical distribution function (EDF) for each marginal and apply the Gaussian copula for the dependence structure. More sophisticated copulas could be selected but we only apply the Gaussian copula for consistency with the other models employed. 	We use  EDF$_n^C$ to denote the empirical marginal distribution with Gaussian copula, calibrated to a sample size $n$. 

	The  Factor Quantile (FQ) methodology developed by \cite{Alexander2020} uses quantile regression to capture the marginals. Each random variable in the system is regressed on a set of $m$ common factors, which may be exogenously selected or determined endogenously. Here we use the latter approach, with latent factors $\mathbf{x}_m$  derived from a principal component analysis of all the variables in the system. Given a quantile partition $\mathcal{Q}$, common to each variable, we perform a set of quantile regressions for each variable, i.e. one for each quantile $\tau \in \mathcal{Q}$. Then, conditional on a set of predetermined values (one for each of the latent factors) we compute the fitted quantiles of each variable, and apply a shape-preserving spline to the fitted quantiles over all $\tau \in \mathcal{Q}$. This way we interpolate a conditional marginal distribution for each variable, all conditional on the same latent-factor values. 	\cite{Alexander2020} claim that such latent-factor FQ models are at least as successful for predicting distributions of financial asset returns as multivariate GARCH models. Furthermore, they have several advantages in that they are very quick to calibrate, more computationally robust than  GARCH models, and they scale far more easily to high dimensions. 
	
	Here we confine our selection to two types of latent factor FQ models, each with a Gaussian copula capturing dependence, which are designated FQ-AL$_{n}^{C}$ and FQ-AB$_{n}^{C}$ where $n$ is again the number of observations used to calibrate the model and $C$ denotes the Gaussian copula.   The difference between the AL and AB versions stems from the choice of factors $\mathbf{x}_m$: the AB version  uses the first $m$ components and the AL version uses the last $m$ principal components. The components are ordered by size of eigenvalue, so that the variance explained by each component progressively decreases.  The AL version simply separates relevant information (captured by the intercept) from the noise (captured by the last $m$ components). In effect, it is a means of removing unwanted variation from the EDF. This may also be viewed as a latent-factor version of the \citet{jensen1968} `alpha',  hence the terminology `alpha-latent' or AL. The AB version uses the first $m$ eigenvectors as latent factors, but in this case each quantile forecast is associated with a large uncertainty.  Therefore a variance reduction technique based on ``bootstrap aggregation'' or \textit{bagging} is applied to reduce this uncertainty,  via an algorithm proposed by \citet{breiman1996}. We refer to \cite{Alexander2020} for further technical details.

\subsection{Dynamic Models}\label{sec:GARCH}
 The most common class of dynamic model for an expected value is the family of autoregressive moving average (ARMA) time-series models --  because they are flexible, have the temporal aggregation property, and allow both easy estimation and straightforward inference. However, for distribution forecasts we also require a parametric forecast of variance. To this end, we employ time-series models for conditional variance forecasts using the  GARCH$(p,q)$ model class, a generalization by \citet{bollerslev1986} of the
autoregressive conditional heteroscedasticity (ARCH) model of \citet{engle1982}. The GARCH$(p,q)$
process assumes a time-varying conditional variance as follows:
\begin{align}
\sigma_{t}^{2} & =\omega+\sum_{i=1}^{q}\alpha_{i}\varepsilon_{t-i}^{2}+\sum_{i=1}^{p}\beta_{i}\sigma_{t-i}^{2}\label{Equation:VanillaGARCHVola}\\
\varepsilon_{t}|\mathcal{I}_{t-1} & \sim\mathcal{N}(0,\sigma_{t}^{2}),\label{Equation:MarketShockDistribution}
\end{align}
where $\varepsilon_{t}$ is the market shock or innovation at time
$t$ and $\mathcal{I}_{t}$ is the information set containing all
past returns up to $t$. The parameters $(\alpha_{i})_{i=1}^{q}$
and $(\beta_{i})_{i=1}^{p}$ measure the reaction of the conditional
variance to market shocks and the persistence of conditional variance
respectively.

These models have
been particularly successful in modelling univariate time-series of financial returns \citep{engle2001}
partly because they are designed to capture volatility clustering effects
which are often present in finance \citep{mandelbrot1963}. In the
presence of such effects, volatility becomes time-dependent and can exhibit prolonged 
periods of exceptionally high or low values.\footnote{GARCH is usually conducive to a two-step maximum likelihood estimation with the first step being the application of an ARMA model to the conditional means and the second being the estimation of a GARCH model to the residuals of the conditional mean equations.} Multivariate GARCH models are designed to capture time-variation in conditional variances and conditional covariances. We only consider the GARCH(1,1) structure. Further lags could be considered, of course, but we need to limit the number of models selected in order to focus our results on the performance of proper scoring rules, and not on the models themselves. Having said this, we should also select the best models in the GARCH$(1,1)$ class, so we shall  allow for asymmetries in innovations as well as leverage effects, and for two kinds of conditional covariance dynamics. In multivariate analysis, clustering extends beyond volatilities to
correlations and generalizations of univariate GARCH models
also capture time-varying conditional covariances and spillover
of volatility between different assets. 
\citet{bollerslev1990} introduces the constant conditional correlation
GARCH (CCC-GARCH) model where the conditional correlations are assumed to
be time-invariant. The covariance matrix is estimated as 
\begin{align*}
\bm{\epsilon_{t}} & =\mathbf{V}_{t}^{1/2}\mathbf{z}_{t}\\
\mathbf{V}_{t} & =\mathbf{D}_{t}\mathbf{C}\mathbf{D}_{t},\\
\mathbf{D}_{t}&=\mbox{diag}\left(\mathbf{V}_{t}\right)^{1/2},
\end{align*}
where $\mathbf{V}_{t}^{1/2}$ is a matrix such that $\mathbf{V}_{t}$
is the covariance matrix (e.g. $\mathbf{V}_{t}^{1/2}$ can be chosen as the Cholesky factorization
of $\mathbf{V}_{t}$)  $\mathbf{z}_{t}$ is the standardised residual which has mean zero
and the identity covariance matrix, $\mathbf{C}$ is a constant correlation
matrix and $\mathbf{D}_{t}$ is the diagonal matrix containing the
time-varying individual volatilities. In practice, each volatility
can be estimated by an univariate GARCH model while $\mathbf{C}$
can be specified as the sample correlation between standardized residuals.
The model is easy to estimate since dependency and the volatilities
are examined separately. 

The assumption of constant correlation may seem too strong and is
not fulfilled for many assets \citep{tsui1999}. Dynamic conditional
correlation GARCH (DCC-GARCH) by \citet{engle2002} generalizes CCC-GARCH
to account for time-varying correlations. The correlation is estimated
directly from the residuals of the univariate models and adjusted
depending on the co-movement of the returns. As such, the covariance
matrix is given by 
\begin{align*}
\mathbf{V}_{t}=\mathbf{D}_{t}\mathbf{C}_{t}\mathbf{D}_{t},\qquad\mathbf{D}_{t}=\mbox{diag}\left(\mathbf{V}_{t}\right)^{1/2},
\end{align*}
where the conditional correlation $\mathbf{C}_{t}$ is described by
\begin{equation}
\begin{aligned}\mathbf{C}_{t} & =\mbox{diag}\left(\mathbf{Q}_{t}\right)^{-1/2}\,\mathbf{Q}_{t}\,\mbox{diag}\left(\mathbf{Q}_{t}\right)^{-1/2},\\
\mathbf{Q}_{t} & =\left(1-\sum_{m=1}^{M}\alpha_{m}-\sum_{n=1}^{N}\beta_{n}\right)\overline{\mathbf{Q}}+\sum_{m=1}^{M}\alpha_{m}\left(\bm{\varepsilon}_{t-m}\bm{\varepsilon}_{t-m}'\right)+\sum_{n=1}^{N}\beta_{n}\mathbf{Q}_{t-n},
\end{aligned}
\label{Equation:DCCCovariance}
\end{equation}
with 
\begin{align*}
\overline{\mathbf{Q}}=\e\left(\bm{\varepsilon}_{t}\bm{\varepsilon}_{t}'\right).
\end{align*}
The transformation of $\mathbf{Q}_{t}$ to $\mathbf{C}_{t}$ guarantees
a well-defined correlation matrix as long as $\mathbf{Q}_{t}$ is positive
definite. Similar to CCC-GARCH, there are no restrictions on the choice
of univariate GARCH models for the volatility. For both the CCC-GARCH and DCC-GARCH models, $\mathbf{z}_{t}$ can
be chosen as the multivariate Gaussian distribution, which leads to
consistent parameter estimators even under distributional misspecification
\citep{bauwens2005}. It is also possible to consider non-normal distributions.
For example, \citet{bauwens2005} use multivariate skew distributions
while \citet{cajigas2006} and \citet{pelagatti2004} apply the Laplace
distribution and general elliptical distributions, respectively. There are several alternative multivariate GARCH models 
but in this paper we limit our analysis to CCC-GARCH(1,1)
and DCC-GARCH(1,1).

\section{Simulation Study}\label{sec:Simulation}

Our discussion in Section 2 introduced several univariate and multivariate
proper scoring rules that are prevalent in the literature. While it
is agreed upon that these offer a sound way to quantify the accuracy
of probabilistic forecasts \citep{winkler1996,gneiting2011}, the
question of which score to use remains largely open \citep{gneiting2007b}.
Conventional wisdom dictates to apply a suitable scoring rule for
the application at hand \citep{machete2013} but this only provides
a few requirements and does not sufficiently restrict the selection.\footnote{As pointed out in Section~2, scoring rules have varying assumptions
for propriety and compare different forecasting types, e.g. density
forecast, distribution forecasts or ensemble forecast, that are sometimes
not easily interchangeable.}

Our aim is to analyse the ability of different scoring
rules to discriminate between true and misspecified distributions. Improving on previous studies 
we employ a realistic simulation setting that
approximates the conditions of practical applications in finance. This
is reflected by our simulation design which calibrates the models described in the previous section to daily USD-denominated
exchange rates from 1999 -- 2018; US interest rates from 1994 --
2018; and Bloomberg investable commodity indices from 1991 -- 2018.

In the following: Subsection \ref{sec:Data} describes the data we
use to calibrate  the models used in the simulation study.  Subsection \ref{sec:ModelSelect} motivates our choice of models.  Subsection \ref{sec:Simulation-Design} describes the design of the
 simulation study.  All results are
discussed in Section \ref{sec:Results}. 

\subsection{Data Description}\label{sec:Data}

We obtain three eight-dimensional time series of daily frequency: the first is on USD-denominated
exchange rates, the second is on US interest rates and the third is on Bloomberg investable commodity
indices. We obtain the daily exchange rates and
commodity index values from Thomson Reuters Datastream and the interest
rates data from the US Treasury website. All time series end on 30
June 2018 but the start date varies with data availability. Within
each set we have selected variables to broadly represent the asset
class. The exchange rates are those with the highest trading volume
excluding the Chinese Renminbi, which was pegged to the USD until recently
\citep{triennial2016}. Our data starts in January 1999 with the introduction
of the Euro as accounting currency. The interest rates span the term
structure of US Treasury bonds from 6 months to 20 years. Alternative
available maturities are 1 month, 2 month, 3 month, and 30 years but
those maturities are missing data for an extended period of time and are therefore excluded.
Our data starts in January 1994 after the 20-year maturity interest
rate became available in October 1993. The commodity indices are
chosen to reflect the most liquid commodities with the highest USD-weighted
production value and are diversified to represent the energy, grains,
industrial / precious metals, softs and livestock sectors \citep{bloomberg2017}.
The Bloomberg commodity indices were launched in 1998 with a backward
projection to January 1991. We include all available data in our study.
We summarize the total sample period and the starting date of our
out-of sample evaluation in Table \ref{Table:DataDescription}.

\begin{table}[htp]
\begin{centering} {\small{}\caption{Sample for each data set}
\label{Table:DataDescription} }%
\begin{tabular}{llll}
\toprule 
{\small{}Data set} & {\small{}First date} & {\small{}Start evaluation} & {\small{}End date}\tabularnewline
\midrule 
{\small{}Exchange rate returns} & {\small{}01 January 1999} & {\small{}28 February 2007} & {\small{}30 June 2018}\tabularnewline
{\small{}Interest rate changes} & {\small{}01 January 1994} & {\small{}03 July 2002} & {\small{}30 June 2018}\tabularnewline
{\small{}Commodity index returns} & {\small{}01 January 1991} & {\small{}26 February 1999} & {\small{}30 June 2018}\tabularnewline
\bottomrule
\end{tabular}{\small{}\medskip{}
}{\small\par}
\end{centering}
{\footnotesize{}All three data sets use daily frequencies, yielding
over 18,000 observations in total. The first
dates vary due to data availability.}{\footnotesize\par}
\end{table}

Summary statistics of the data are listed in Table \ref{Table:SummaryStatistics} for monthly returns of exchange rates and commodites, and for monthly changes in interest rates. The sample mean for all monthly
returns and changes is near zero which allows us to apply the principal
component representation of Factor Quantile models without prior transformations.
Furthermore, all assets are leptokurtic and require heavy-tailed distributions.

\begin{table}[h]
\begin{centering}
{\small{}\caption{Summary statistics of the monthly returns / changes}
\label{Table:SummaryStatistics} }%
\begin{tabular}{lllll}
\toprule 
{\small{}Asset} & {\small{}Mean} & {\small{}Volatility} & {\small{}Skewness} & {\small{}Kurtosis}\tabularnewline
\midrule 
\multicolumn{4}{l}{{\small{}Exchange rate returns}} & \tabularnewline
{\small{}AUD} & {\small{}0.0000} & {\small{}0.0368} & {\small{}0.7922} & {\small{}5.9807}\tabularnewline
{\small{}CAD} & {\small{}-0.0003} & {\small{}0.0265} & {\small{}0.7218} & {\small{}6.4690}\tabularnewline
{\small{}CHF} & {\small{}-0.0011} & {\small{}0.0297} & {\small{}-0.0101} & {\small{}4.8305}\tabularnewline
{\small{}EUR} & {\small{}0.0003} & {\small{}0.0292} & {\small{}0.3192} & {\small{}4.1236}\tabularnewline
{\small{}GBP} & {\small{}0.0013} & {\small{}0.0252} & {\small{}0.5218} & {\small{}4.9380}\tabularnewline
{\small{}JPY} & {\small{}0.0002} & {\small{}0.0281} & {\small{}0.3077} & {\small{}3.5187}\tabularnewline
{\small{}NZD} & {\small{}-0.0003} & {\small{}0.0383} & {\small{}0.5727} & {\small{}4.6172}\tabularnewline
{\small{}SEK} & {\small{}0.0011} & {\small{}0.0328} & {\small{}0.1555} & {\small{}3.5244}\tabularnewline
 &  &  &  & \tabularnewline
{\small{}6 month} & {\small{}-0.0039} & {\small{}0.2016} & {\small{}-2.2169} & {\small{}14.6083}\tabularnewline
{\small{}1 year} & {\small{}-0.0041} & {\small{}0.2137} & {\small{}-1.2097} & {\small{}8.6316}\tabularnewline
{\small{}2 year} & {\small{}-0.0055} & {\small{}0.2462} & {\small{}-0.3439} & {\small{}4.4936}\tabularnewline
{\small{}3 year} & {\small{}-0.0062} & {\small{}0.2617} & {\small{}-0.0753} & {\small{}3.9820}\tabularnewline
{\small{}5 year} & {\small{}-0.0078} & {\small{}0.2728} & {\small{}0.0288} & {\small{}3.7443}\tabularnewline
{\small{}7 year} & {\small{}-0.0086} & {\small{}0.2697} & {\small{}0.1118} & {\small{}3.7856}\tabularnewline
{\small{}10 year} & {\small{}-0.0097} & {\small{}0.2594} & {\small{}-0.0196} & {\small{}4.2279}\tabularnewline
{\small{}20 year} & {\small{}-0.0115} & {\small{}0.2372} & {\small{}0.0341} & {\small{}4.6910}\tabularnewline
 &  &  &  & \tabularnewline
{\small{}Copper} & {\small{}0.0065} & {\small{}0.0724} & {\small{}-0.0517} & {\small{}5.8856}\tabularnewline
{\small{}Corn} & {\small{}-0.0048} & {\small{}0.0752} & {\small{}0.2992} & {\small{}4.0710}\tabularnewline
{\small{}Gold} & {\small{}0.0027} & {\small{}0.0453} & {\small{}0.1885} & {\small{}4.1817}\tabularnewline
{\small{}Live Cattle} & {\small{}-0.0006} & {\small{}0.0392} & {\small{}-0.4110} & {\small{}5.1238}\tabularnewline
{\small{}Natural Gas} & {\small{}-0.0080} & {\small{}0.1316} & {\small{}0.4827} & {\small{}3.9878}\tabularnewline
{\small{}Soybean} & {\small{}0.0047} & {\small{}0.0684} & {\small{}-0.0485} & {\small{}3.5828}\tabularnewline
{\small{}Sugar} & {\small{}0.0037} & {\small{}0.0886} & {\small{}0.2235} & {\small{}3.4728}\tabularnewline
{\small{}WTI Oil} & {\small{}0.0034} & {\small{}0.0876} & {\small{}-0.0136} & {\small{}3.8353}\tabularnewline
\bottomrule
\end{tabular}{\small{}\medskip{}
}{\small\par}
\par\end{centering}
{\footnotesize{}The monthly returns and changes are calculated using
the values at the start of each month which the summary statistic
aggregates over the time periods specified in Table~\ref{Table:DataDescription}.
Our study only applies daily data but we use a monthly frequency in
this table to avoid minuscule magnitudes.}{\footnotesize\par}
\end{table}
To examine the robustness of our analysis, we segment the data into
three parts, covering 2006--2010, 2010--2014, and 2014--2018
with breakpoints at the end of June in each case. Because the exchange
rate data starts much later than the other data sets, the period from
June 2006 to February 2007 is still used for calibration. Therefore,
the first sub-period begins in March 2007 in this case.

\subsection{Model Selection}\label{sec:ModelSelect}

Although there have been a few studies explicitly
 	comparing scoring rules (as reviewed in Section~\ref{sec:Introduction}), the models chosen and the corresponding discussions generally fail to put enough emphasise on the distinct features of financial data, such
 	as time-varying volatility, covariance and leptokurtic conditional
 	distributions. \citet{pinson2013} restrict themselves to bivariate Gaussian distributions with different means, variances and correlations, and do not consider time series in
 	their study. Similarly \citet{scheuerer2015} only consider simple distributions
 	such as Gaussian and Poisson distributions.\footnote{
		This comparison of multivariate scoring rules is more comprehensive
		than alternative ones, but still mainly uses Gaussian distributions
		as the DGP. In the case where a Poisson distribution is assumed as
		the DGP, all scores but the variogram score with $p=0.5$ have at least
		some ranking issues and may identify the wrong model as the correct
		one.}  \citet{ziel2019multivariate}
 	conduct a time series empirical study but they only consider AR models
 	for the mean with no dynamic volatility structure. In addition, their
 	data is essentially a univariate time series, where observations on multiple days are viewed jointly as multivariate distributions. This is inherently different from standard financial applications, where synchronised
 	financial data are viewed as a multivariate system. \citet{Diks2020531}
 	do consider financial applications, in which they use the univariate
 	and multivariate GARCH models to capture the dynamic variance and
 	covariance structure. However, as we explained in Section~\ref{sec:Introduction},
 	their focus is not to compare scoring rules.
 	
    Most of the empirical literature on financial forecasting accuracy concerns univariate times series and, moreover, considers only point forecasts. Yet, even in this restricted context, it remains unclear whether a dynamic model structure is always preferable to a static one which assumes that the next period distribution of variables can be well approximated by the historical distribution.  Furthermore, multivariate models can differ both according to the specification of the marginals and their dependence structure. As described in Section \ref{sec:Models}, a static marginal distribution is commonly represented by an empirical distribution function, while we also consider models in the Factor Quantile class. The dynamic models are instead from the GARCH class. Regarding dependence, all but one of our models assumes a static dependence structure (typically the Gaussian copula) -- only the DCC-GARCH models extends this assumption to dynamic conditional correlation.  
    
	Our simulation study selects a DGP as the true distribution and assesses the ability of different proper scoring rules to detect model misspecification. To reduce dependence on our choice of DGP we employ a diverse selection of models for multivariate distributions that can be used as either the DGP or a misspecified model. Also, for our results to be relevant to practical  applications we calibrate these models to a very comprehensive selection of real-world data. However, even though the models are calibrated to actual data, it is important to emphasize that we do not seek to compare the forecasting performance of different models. Instead, our focus is on a comparison of the ability of different proper scoring rules to detect model misspecification.  Nonetheless, we do choose models which have been shown in \cite{Alexander2020} to have similar forecasting abilities in typical financial datasets, thereby posing a greater challenge to scoring rules in terms of identifying the DGP and discriminating between competing models.  As compared with other studies, we thus emply a more realistic setting with models that are better representations of financial data, we calibrate these models to actual data and our results are much more comprehensive.

We employ a sample size $n=250$ and $n=2000$ for our results based on static models. However, the dynamic multivariate GARCH models cannot be calibrated using $n=250$  because the optimization of their complex likelihood function does not converge, most of the time. Hence, their parameters are estimated using $n=2000$ only.  The FQ models are much easier to calibrate on small samples, so we present results for these models based on both $n=250$ and $n=2000$.\footnote{Results for other choices of $n$ are available, as are those with the assumption of independent marginals. However, these are excluded here for brevity and because their use does not change our conclusions about the ability of proper scoring rules to discriminate misspecified models -- but they are available from the authors on request.} Because the EDF and FQ models are calibrated on two different sample sizes we end up with eight
competing models in total: 
(i) Two FQ-AL models, (ii) two FQ-AB models, (iii)
two EDF models and (iv) two multivariate GARCH models. This way,
each model has one associated model that is similar but differs either
in the calibration length or the conditional covariance structure.  Table \ref{Table:ModelDescription} summarizes the models employed.

\begin{table}[!htb]
\centering {\small{}\caption{Summary of Models used in the Simulation Study}
\label{Table:ModelDescription} }%
\begin{tabular}{lllr}
\toprule 
{\small{}Model} & {\small{}Marginals} & {\small{}Dependency} & {\small{}Calibration}\tabularnewline
\midrule 
{\small{}$\mbox{EDF}_{250}^{C}$} & {\small{}EDF} & {\small{}Gaussian copula} & {\small{}250}\tabularnewline
{\small{}$\mbox{EDF}_{2000}^{C}$} & {\small{}EDF} & {\small{}Gaussian copula} & {\small{}2,000}\tabularnewline
{\small{}$\mbox{FQ-AL}_{250}^{C}$} & {\small{}Alpha FQ w/ last PC} & {\small{}Gaussian copula} & {\small{}250}\tabularnewline
{\small{}$\mbox{FQ-AL}_{2000}^{C}$} & {\small{}Alpha FQ w/ last PC} & {\small{}Gaussian copula} & {\small{}2,000}\tabularnewline
{\small{}$\mbox{FQ-AB}_{250}^{C}$} & {\small{}Asym. Bagging FQ} & {\small{}Gaussian copula} & {\small{}250}\tabularnewline
{\small{}$\mbox{FQ-AB}_{2000}^{C}$} & {\small{}Asym. Bagging FQ} & {\small{}Gaussian copula} & {\small{}2,000}\tabularnewline

{\small{}CCC-GARCH} & {\small{}Student-t E-GARCH(1,1)} & {\small{}Conditional correlation} & {\small{}2,000}\tabularnewline
{\small{}DCC-GARCH} & {\small{}Student-t E-GARCH(1,1)} & {\small{}Dyn. cond. correlation} & {\small{}2,000}\tabularnewline
\bottomrule
\end{tabular}{\small{}\medskip{}
}{\small\par}
\raggedright{}{\footnotesize{}Column 1 summarizes the notation used for the model, column 2 describes the marginals, column 3 the dependency structure and column 4 the length of the calibration window. For example, the GARCH models are calibrated to an eight-dimensional  time series spanning 2000 days and the FQ-AB$_{250}^C$ models are calibrated to an eight-dimensional  time series spanning 250 days.}{\footnotesize\par}
\end{table}
	
		 Our motivation for selecting the models described here is to use a small but representative selection from the static and dynamic modelling paradigms that are commonly used for multivariate financial systems. Expanding the model set beyond the eight models described in this section may hinder rather than help our aim.  Our selection of models strikes a balance between the need to represent the salient qualities of real data adequately and a sufficiently simple structure that calibration difficulties will not interfere with the very large scale of our simulation study.

\subsection{Simulation Design\label{sec:Simulation-Design}}

Our simulation study quantifies and compares the ability of the energy score and
the variogram score with $p=0.5,1,2$ to distinguish the correct DPG
from misspecified models. These scoring rules were defined in Section \ref{sec:Rules} and the selected values of $p$, which have also been used by \citet{scheuerer2015} are considered typical choices \citep{jordan2017}.  

The models described above are calibrated
on the systems of daily, eight-dimensional USD-denominated exchange rates,
interest rates and Bloomberg investable commodity indices that we
discussed in Section \ref{sec:Data}.  
The parameters for each model are estimated on an initial sample size $n$, then the calibrated model is used to generate a one-day-ahead multivariate distribution forecast, then the sample is rolled forward keeping the sample size fixed, and then the calibration and forecasting is repeated until the data set is exhausted. Then we move to another data set and repeat, until we have a continuous set of daily, rolling distribution forecasts, for each model and for each of the three data sets. 
	
Our aim is not to assess the forecasting performance of different models. It is to assess the ability of a scoring rule to distinguish 
	the true DGP from misspecified models, when used for multivariate distribution forecasting.   Our simulation setting controls the DGP so that we
know the true distribution every time we evaluate the forecasts and the other seven models are misspecified.   For the forecasts made at time $t$ all the models are calibrated using the $n$ most recent observations up to time $t$, as previously mentioned. In addition, we select one of these models to generate the observations for time $t+1$ that are used to analyse the performance of scoring rules. Put another way, realisations are simulated from just one of these models -- i.e. the one that is designated as the  DGP. The reason why we have selected such diverse models is that we seek to reduce
the dependence of our results on the choice of a specific DGP. Therefore, we rotate the choice of DGP to span all eight  
models.  

The simulation algorithm is summarized as follows. First, fix a proper scoring rule $s$ and a model $m^*$ that is designated as the true DGP. This rule $s$ will quantify each of the eight model's  distribution forecasts, for each data set, as follows: 
\begin{description}
\item [{Stage 1}] Calibrate
	a model  using data ending at time $t$ and forecast a multivariate  distribution
for time $t+1$, i.e. one-day-ahead. Repeat this eight times, once for each model in Table \ref{Table:ModelDescription};
\item [{Stage 2}] Draw 5,000 observations from the  distribution forecasted by
model $m^*$. These observations are realisations from the true model at time $t+1$; 
\item [{Stage 3}] Given a distribution forecast for time $t+1$ by model $m$, apply $s$ to each of the 5,000 realisations generated in Stage 2, to quantify the performance of 
 model $m$. Repeating this for all models in Table \ref{Table:ModelDescription} gives  5,000 scores for each of the eight models at time $t$ including the designated DGP model $m^*$;
\item [{Stage 4}] Roll $t$ forward by one quarter of a year and repeat stages 1, 2 and 3 until the entire sample for that data set is exhausted.
\end{description}
The four stages outlined above are then repeated eight times, so that each of the models in Table \ref{Table:ModelDescription} is selected as the designated DGP $m^*$. Then we change the scoring rule $s$ and repeat the four stages eight times, so that again each model can be the designated DGP $m^*$. We do this four times, so that $s$ spans all four scoring rules and every time we change the scoring rule we repeat stages 1 to 4 eight times, so that  each of the eight models can be the designated  DGP. Finally, we change the data set and repeat the entire process again.

We evaluate the scoring rules at the first date of each quarter in
our evaluation period. This yields new simulations (each with eight dimensions) on 50, 66, and 82 dates
in USD-denominated exchange rates, US interest rates and Bloomberg
investable commodity indices, respectively. Since
we have eight possible DGPs, this leads to approximately 1,600 applications
of the simulation study above for each of the four multivariate scoring
rules. 

This setting gives us a very detailed view on the discrimination
ability for each scoring rule over time (i.e. different market conditions) and for various choices of
the DGP. Note that our simulation design provides optimistic conditions for the scoring
rules since it knows the distribution of the DGP for certain, and moreover it samples a very
large number of realisations at each time $t$.\footnote{Indeed, each simulation yields
	 a scoring-rule evaluation based on observations with a stationary DGP.  Depending on the realisation, the scoring rules may favour a model other than the DGP but the sample mean based on all 5,000 scores should
	be the smallest for the DGP. This, of course, is because the distribution
	of the DGP is used to generate the realisations. A good scoring rule
	should assign the lowest scores to the DGP and also produce robust
	rankings over the entire evaluation period. As \citet{pinson2013}
	point out, a large distance between the scores of the DGP and alternative
	models may help to avoid erroneous conclusions.}  In practice, we only
observe one realisation and therefore must consider the scores over
a large period -- and the longer the sample used the less likely the DGP is to be stationary. To approximate a more realistic setting, in which the length of the
out-of-sample period is restricted due to lack of data, we shall also compare
the scoring rules on smaller sub-samples, with only 100 realisations instead of 5,000. Recall that all
models use historical information to forecast their joint distribution, but in our setting
they are evaluated using samples from the chosen DGP, rather
than realisations of the original time series. The models are only
punished if their forecast deviates from that of the true DGP and a good scoring
rule should be able to distinguish misspecified  models from the
true distribution.  

\subsection{Discrimination Metrics}\label{sec:Metrics}
 We shall generalize
the approach of \citet{pinson2013} to compare the discrimination
ability of several scoring rules and introduce a new metric, the error rate, as an
additional measure of the sensitivity of scoring rules. Suppose there are $M$ models in the study and that scores are based on $N$ realisations at  time $t$. The $N \times 1$ score vector assigned by scoring rule $s$ to model $m$ at time $t$, when model $m^{*}$ is the DGP is defined as:
\begin{align}
\mathbf{S}_{t}^{s}(m,m^{*})=\left(S_{1,t}^{s}(m,m^{*}),\ldots,S_{N,t}^{s}(m,m^{*})\right)'.
\label{Equation:Score}
\end{align}
We employ several metrics aimed at summarizing the properties of \eqref{Equation:Score}. First, and following \cite{scheuerer2015}, 
the simplest  metric we employ is the mean relative score:
 \begin{align}
\frac{1}{N}\sum_{i=1}^{N}\left(S_{it}^{s}\left(m,m^{*}\right)/S_{it}^{s}\left(m^{*},m^{*}\right)\right).\label{Equation:RelativeSampleMean}
\end{align}
However, by construction, the mean relative score may yield
erroneous rankings when $N$ is small. For some realisations, the
lowest score may be assigned to a model that is not the DGP. We study
this probability in our simulation study by introducing a new `error rate'
metric and also by analysing the distribution of the absolute differences between the scores assigned
to each model:
\begin{align}
\mathbf{S}_{t}^{s}\left(m,m^{*}\right)-\mathbf{S}_{t}^{s}\left(m^{*},m^{*}\right).\label{Equation:AbsoluteErrorScoringRules}
\end{align}
As an additional measure for the discrimination ability, we consider
a simple heuristic that examines the relative distance of the scores
between the models. This approach has been suggested by \citet{pinson2013}
who compare the sensitivity of the energy score to various misspecifications.
Given 
\begin{align*}
\overline{\mathbf{S}}_{t}^{s}\left(m,m^{*}\right)\defeq\frac{1}{N}\sum_{i=1}^{N}S_{it}^{s}\left(m,m^{*}\right),
\end{align*}
they utilize a true Gaussian distribution and measure the sensitivity pairwise through
\begin{align*}
\frac{\overline{\mathbf{S}}_{t}^{s}\left(m,m^{*}\right)-\overline{\mathbf{S}}_{t}^{s}\left(m^{*},m^{*}\right)}{\overline{\mathbf{S}}_{t}^{s}\left(m^{*},m^{*}\right)}
\end{align*}
with misspecified models that deviate from the true distribution in only in one aspect, e.g. mean or
variance. For our purposes we need to  adjust their measure to
consider the discrimination across scoring rules over multiple misspecified DGPs. To this end, we propose a generalized discrimination heuristic
that is defined as 
\begin{align}
d_{t}^{s}(m^{*})=\frac{1}{M}\sum_{m=1}^{M}\frac{\overline{\mathbf{S}}_{t}^{s}\left(m,m^{*}\right)}{\overline{\mathbf{S}}_{t}^{s}\left(m^{*},m^{*}\right)}.\label{Equation:DisciminationHeuristic}
\end{align}
A large value for this generalised discrimination metric may indicate that the ranking
of the scores is reliable and robust to differences in the simulation sample size $N$.


\section{Simulation Results}\label{sec:Results}

We analyse the energy score and three parametrisations of the variogram
score with respect to their ability to identify the DGP under the
simulation design described in the previous section. As discussed in Section \ref{sec:Simulation-Design}, we use $N=5000$
and $M=8$. To examine the discrimination ability of each scoring
rule at time $t$, we apply the entire sample of 5,000 scores but
also smaller sub-samples with $N = 100$ scores that reflect more realistic
conditions. These correspond to out-of-sample evaluations with 5,000
and 100 observations respectively.
Each score assesses the closeness of the distribution forecasts to the
distribution of the DGP, rather than indicating real forecasting accuracy.  Because we impose a distribution through the DGP, a model's scores only measure how similar the models is to the DGP, according to that score. 

We begin our results in Section \ref{subsec:Sample-Mean-Comparison} with the mean relative score \eqref{Equation:RelativeSampleMean}, focusing on exchange
rate returns and a DCC-GARCH as DGP. 
Then, in Section \ref{subsec:Error-Rate-Comparison} we generalize these results to multiple DGPs and data sets by
using the error rate to measure the percentage of cases in which a misspecified model receives
a lower (i.e. better) score than the DGP. Further, we analyse the
deviation between the scores of misspecified models and that of the
DGP, derived from \eqref{Equation:AbsoluteErrorScoringRules}. Finally, in Section \ref{subsec:Discrimination-Heuristics}, we illustrate the distribution of these
deviations and use our generalization \eqref{Equation:DisciminationHeuristic}  of the discrimination heuristic proposed by
\citet{pinson2013} 
that allows us to compare multiple scoring rules. For reasons of space several figures
are only available electronically in the supplementary materials. There is also an expanded set of results where the model set used includes some where the correlation matrix is replaced with the identity matrix. 

\subsection{Mean Relative Score\label{subsec:Sample-Mean-Comparison}}

\begin{figure}[!htb]
\caption{Average scores relative to score of DGP (USD exchange rates)}
\label{Figure:MeanScores} \centering \includegraphics[width=1\textwidth]{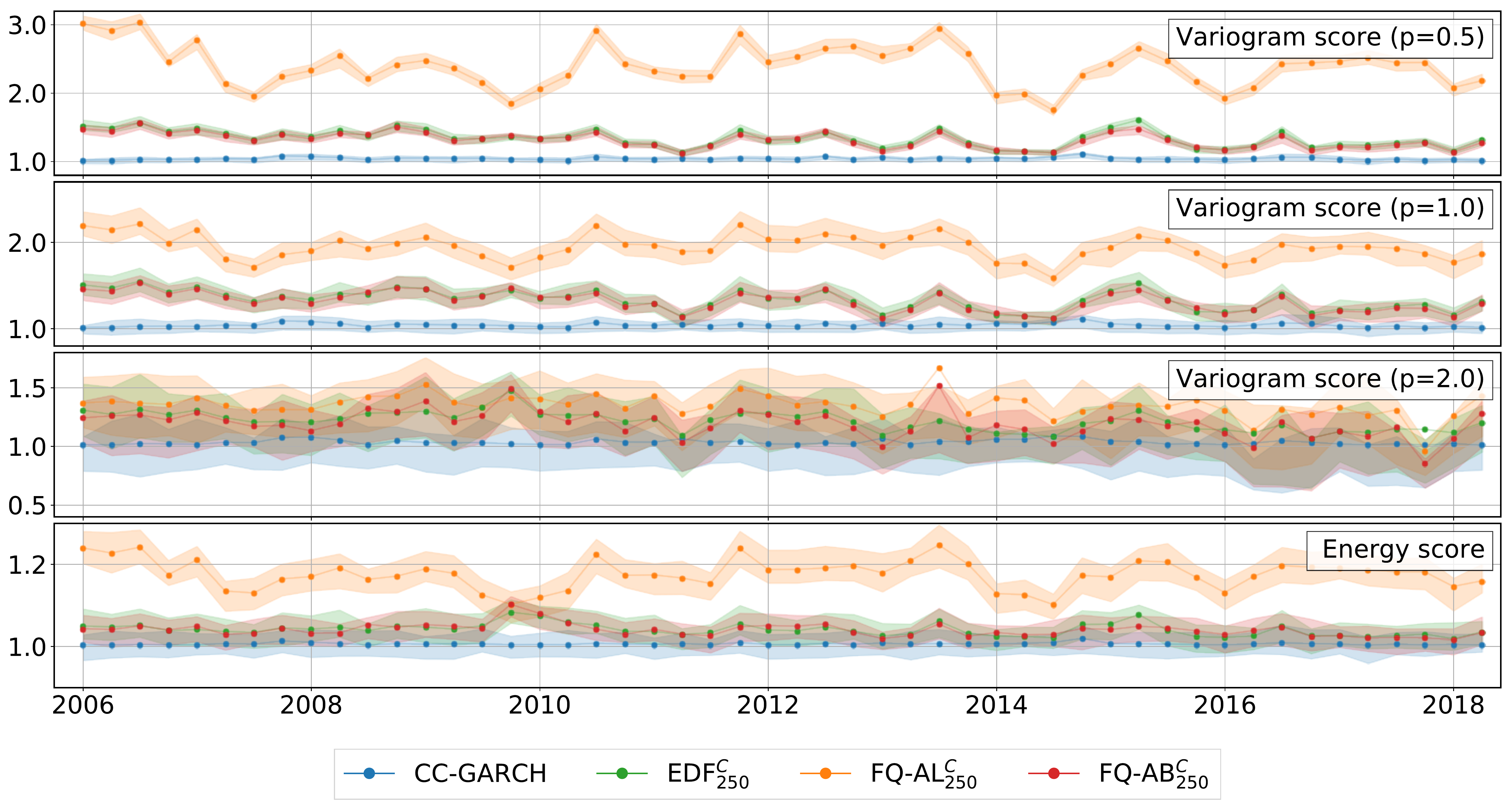} 
\raggedright{}{\footnotesize{}The figure illustrates the 
mean relative score in Equation~\ref{Equation:RelativeSampleMean}
based on 5,000 scores. A value larger than 1 means that the scoring
rule is on average able to distinguish between the misspecified model
and }%
\mbox{%
{\footnotesize{}DCC-GARCH}%
}{\footnotesize{} to identify the true DGP. We generate a confidence
interval covering the area between the 0.25- and 0.75-quantiles of
the sample mean based on a sample size of 100 through bootstrap with
5,000 repetitions.}{\footnotesize\par}
\end{figure}

We begin the analysis of the multivariate scoring rules with a comparison
of their mean relative score \eqref{Equation:RelativeSampleMean} for each model. Figure~\ref{Figure:MeanScores}
uses a %
\mbox{%
DCC-GARCH%
} as DGP for exchange rate returns for four selected models, with the DGP ($m^*$) in this case being DCC-GARCH. The mean relative score corresponds to the average of the ratio 
between the score of the misspecified model and that of the DGP. The shaded areas cover everything between the 0.25- and 0.75-quantiles
for the sample mean based on a sample size of 100 instead of 5,000.
These confidence intervals are generated through a statistical bootstrap
with 5,000 repetitions. We limit the illustration to four models only
for clarity, but the results are comparable when other misspecified
models, DGPs or data sets are considered. Figures for other DGPs and
data sets can be found in the supplementary materials.

The results based on the sample mean of 5,000 scores indicate that
all four scoring rules manage to evaluate the models successfully.
Due to propriety, they assign the lowest expectation to the DGP which
is why almost none of the mean relative scores fall below 1 in Figure~\ref{Figure:MeanScores}.
Further, %
\mbox{%
CCC-GARCH%
} almost always obtains the lowest score among all misspecified models
as expected given its similarity to %
\mbox{%
DCC-GARCH%
}. The scores can distinguish distributions which differ only in their
marginals and they are able to identify %
\mbox{%
FQ-AL$_{250}^{C}$%
} as one of the misspecified models with great confidence. In contrast,
the difference between %
\mbox{%
FQ-AB$_{250}^{C}$%
} and %
\mbox{%
EDF$_{250}^{C}$%
} is less pronounced which means that they produce predictions 
which each deviate by a similar amount to the distribution forecast of %
\mbox{%
DCC-GARCH%
}.

Both the variogram score with $p=0.5$ and $p=1$ show clear and robust
rankings between the misspecified models and distinguish them well from
the DGP. The discrimination ability is weaker for the energy score.
As pointed out by \citet{pinson2013} and \citet{scheuerer2015},
the energy score changes only by a small amount between the DGP and
other models. This is evident in Figure~\ref{Figure:MeanScores}
as well, where the average score of the worst model is only 25\% larger
than that of the DGP. In comparison, the variogram scores with $p=0.5$
and $p=1$ assign average scores over 200\% and 100\% larger than
that of the DGP respectively. Unlike the other scoring rules, the
variogram score with $p=2$ changes the rankings at several times
and is also the only scoring rule which makes wrong inferences even
with a large sample size of 5,000 scores. For instance, %
\mbox{%
FQ-AB$_{250}^{C}$%
} is preferred over the DGP around the end of 2017. Hence, the energy
score and variogram scores with $p=0.5$ and $p=1$ may be preferable
to the variogram score with $p=2$.

However, there are vast differences in the discrimination ability of scoring rules
which can lead to wrong inferences in smaller sample sizes: 
\begin{enumerate}
\item Despite the overall success of the variogram score with $p=0.5$ and
$p=1$, wrong inferences may occur with only 100 samples. The shaded
areas of %
\mbox{%
CCC-GARCH%
} dip below 1 frequently which means that a slightly misspecified model
may be chosen over the DGP. 

\item The energy score suffers from misidentification risk to a much larger extent.
Besides %
\mbox{%
CCC-GARCH%
}, %
\mbox{%
FQ-AB$_{250}^{C}$%
} and %
\mbox{%
EDF$_{250}^{C}$%
} are also at times assigned lower scores than the DGP in 2010, 2013, 2016 and
2017 with the smaller sample size. Overall though, the energy score still manages to produce a
clear ranking that is mostly accurate. 

\item The variogram score with $p=2$ largely fails to yield any meaningful
results with the smaller sample size. The rankings can change considerably,
and all models obtain a lower sample mean than the DGP at various
times. Even %
\mbox{%
FQ-AL$_{250}^{C}$%
}, which is regarded as the worst model by all other scoring rules,
has lower scores than %
\mbox{%
DCC-GARCH%
} around 2016. Additionally, the variogram score with $p=2$ may assign
scores of very large magnitude that greatly affect the sample mean.
This is visible in Figure~\ref{Figure:MeanScores} in two aspects:
(i) The scoring rule has wide confidence intervals and (ii) the sample
mean is at times higher than the sample 0.75-quantile. This is, for
instance, the case around the end of 2013. 
\end{enumerate}
These initial findings suggest that variogram score with $p=0.5$
and $p=1$ offer superior discrimination ability to the more popular
energy score. The variogram score with $p=2$ performs very poorly
and may yield erroneous rankings of the forecasting models, even with
a very large sample of scores.

\subsection{Error Rate Comparison\label{subsec:Error-Rate-Comparison}}

\begin{figure}[p]
\caption{Density of differences between scores with DCC-GARCH as DGP}
\label{Figure:ErrorDistribution} \centering \includegraphics[width=1\textwidth]{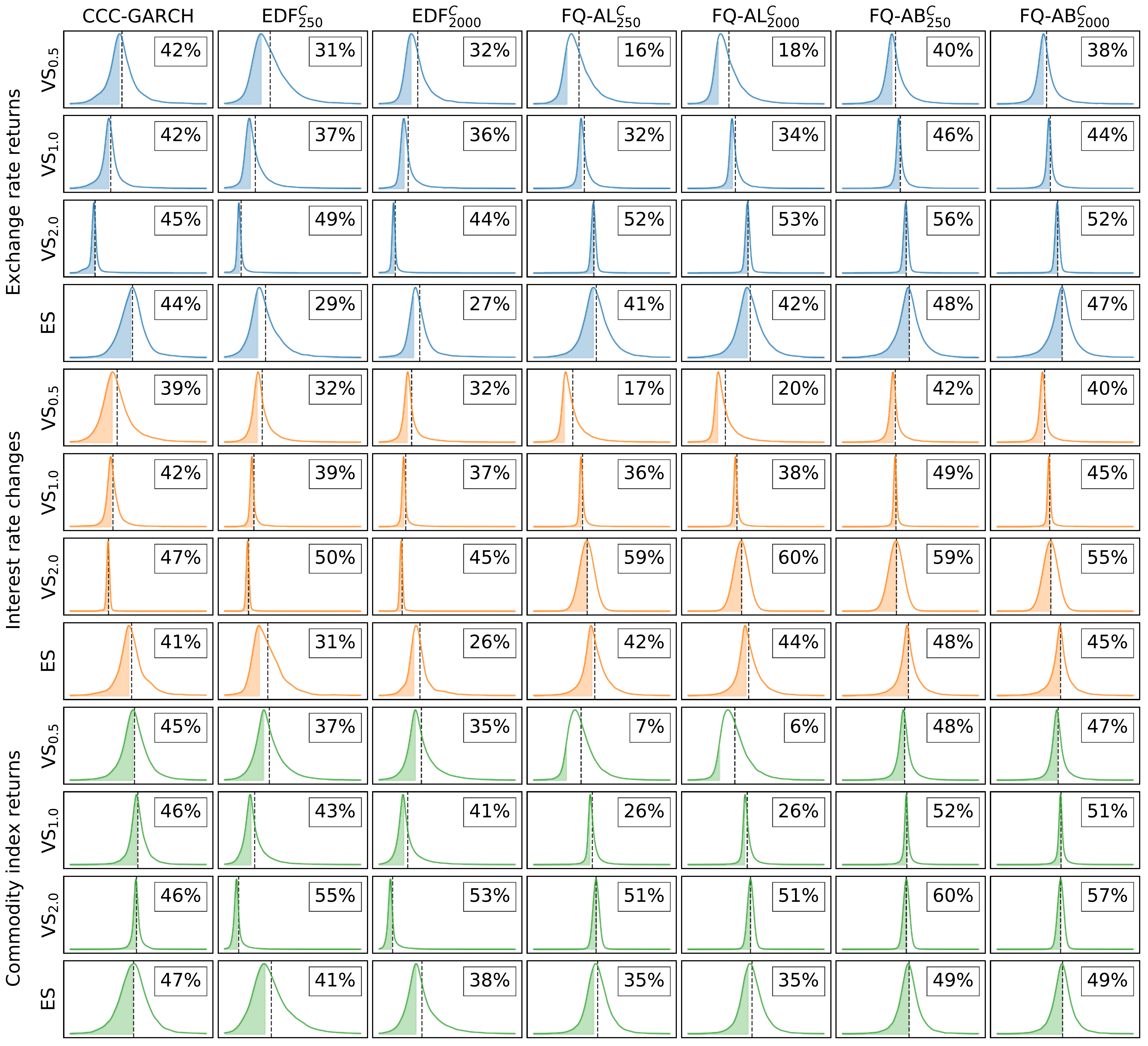} 
\raggedright{}{\footnotesize{}This figure displays the density of
the difference between the scores of the DGP and the misspecified
models described in equation}~(\ref{Equation:AbsoluteErrorScoringRules}){\footnotesize{}.
A Gaussian kernel is used to smooth the densities. The shaded areas
correspond to negative values, where a lower score is assigned to
the misspecified models. In the upper right corner of each sub-figures,
the probability of the shaded area is displayed. The dotted vertical
line shows the expectation of the density. For clarity, we limit the
sub-figure to values between the 0.001- and 0.999-quantiles. Figures
for alternative DGPs can be found in the supplementary materials.}{\footnotesize\par}
\end{figure}

Figure \ref{Figure:ErrorDistribution} shows the results from applying \eqref{Equation:AbsoluteErrorScoringRules} with
\mbox{%
DCC-GARCH%
} as DGP for all datasets and all competing models and uses the scores for all $t$ to generate the density.
Each column of the figure illustrates the density of equation~(\ref{Equation:AbsoluteErrorScoringRules})
for a specific misspecified model, under various scoring rules and
data sets. We include the error rate in the upper right corner of
each sub-figure which shows the probability that equation~(\ref{Equation:AbsoluteErrorScoringRules})
yields a negative value. For clarity, we do not use the same x-axis
for all sub-figures but show all values between the 0.001- and 0.999-quantiles
of each distribution. This means that the magnitude of the error is
not visible in these figures but instead we gain insight on the shape
of the error density. Figures for alternative DGPs can be found in
the supplementary materials.

Overall, Figure \ref{Figure:ErrorDistribution} shows that the probability
of getting scores which are lower than that of the DGP is 
quite high and
varies significantly across cases.  Depending on the data set and scoring
rule, the average error rate (for each row of plots) varies between 31\% and 54\%.  The variogram score with $p=2$ in particular often assigns lower scores to misspecified models. This happens in 50\%, 54\%, 53\% of
cases for exchange rate returns, interest rate changes and commodity
rate returns respectively and is therefore around 60\% worse than
the error rate of the variogram score with $p=0.5$ (with corresponding error rates of 31\%, 32\% and 32\%). This scoring
rule achieves the lowest error rate, followed by the variogram score
with $p=1$ and the energy score, which both achieve error rates in the 39-42\% range.

It is important to note that the error rate is only a binary statistic
which does not take into account the magnitude by which the scores
of misspecified models are smaller than that of the DGP. By averaging
over a sample of scores, the error rate decreases, until it reaches
zero due to the propriety of the scoring rules. The number of samples
needed for a sample mean that favours the DGP depends on the shape
of the distribution. If the tail of the shaded area is small in comparison
to the tail of the non-shaded area, a small sample might be sufficient.
However, many of the distributions in Figure \ref{Figure:ErrorDistribution}
are approximately symmetric which means that large positive and negative
values in equation~(\ref{Equation:AbsoluteErrorScoringRules}) are
equally likely. As an additional indicator for the convergence speed,
we illustrate the expectation of the distributions with a dotted line.
These are always non-negative due to propriety of the scoring rules,
but an expectation far right from the shaded area corresponds to a
faster convergence towards lower mean relative scores for the DGP. Again,
the values are generally close to zero, which suggests
slow convergence towards positive sample mean scores.

\begin{figure}[!htb]
\caption{Error rates of scoring rules}
\label{Figure:ErrorRate} \begin{centering}
 \includegraphics[width=1\textwidth]{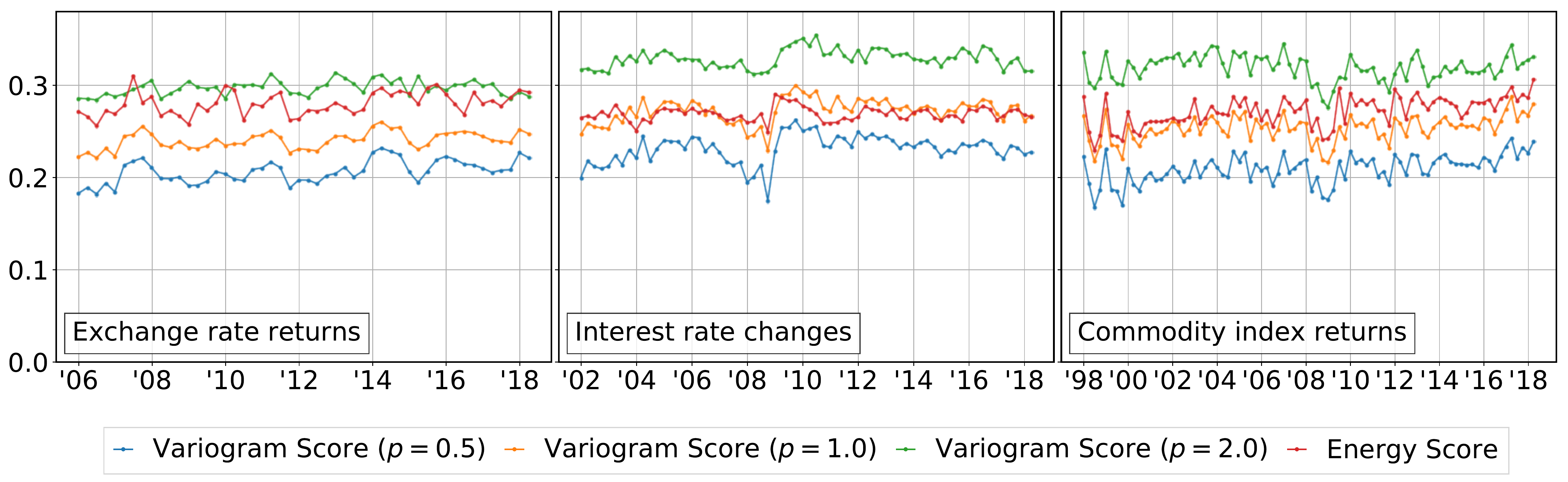}
\end{centering}
\raggedright{}{\footnotesize{}The error rates show how often a misspecified model
is assigned a lower score than the DGP. Higher values are associated
with inferior scoring rules and more frequently wrong inferences.}
\end{figure}

The average error rate over all DGPs for the evaluation period of
the multivariate scoring rules is compared in Figure \ref{Figure:ErrorRate}.
Similar to Figure \ref{Figure:ErrorDistribution}, we examine the
number of times the score of a misspecified model is lower than that
of the DGP but we now include the error rate across multiple choices
of the DGP. The conclusions from Figure \ref{Figure:ErrorRate} are similar to Figure
\ref{Figure:ErrorDistribution}, although the magnitude of the error rates is lower than for the case of only DCC-GARCH as DGP.  The variogram score with $p=2$ has
a significantly higher error rate that is more than 47\% higher than
that of the variogram score with $p=0.5$. This is also consistent
with Figure \ref{Figure:MeanScores} where misspecified models were
sometimes preferred over the DGP. Again, there is typically a clear ranking of the scoring
rules that persists with all three data sets and the entire evaluation
period. For the variogram scores, the error rate increases with the
parameter $p$ and the error rate of the energy score typically falls
between the variogram score with $p=1$ and $p=2$, but interestingly it often outperforms variogram with $p=1$ for the interest rate data.

\subsection{Generalised Discrimination Heuristic\label{subsec:Discrimination-Heuristics}}

\begin{figure}[p]
\caption{Discrimination heuristic of scoring rules}
\label{Figure:Discrimination} \begin{centering}
 \includegraphics[width=1\textwidth]{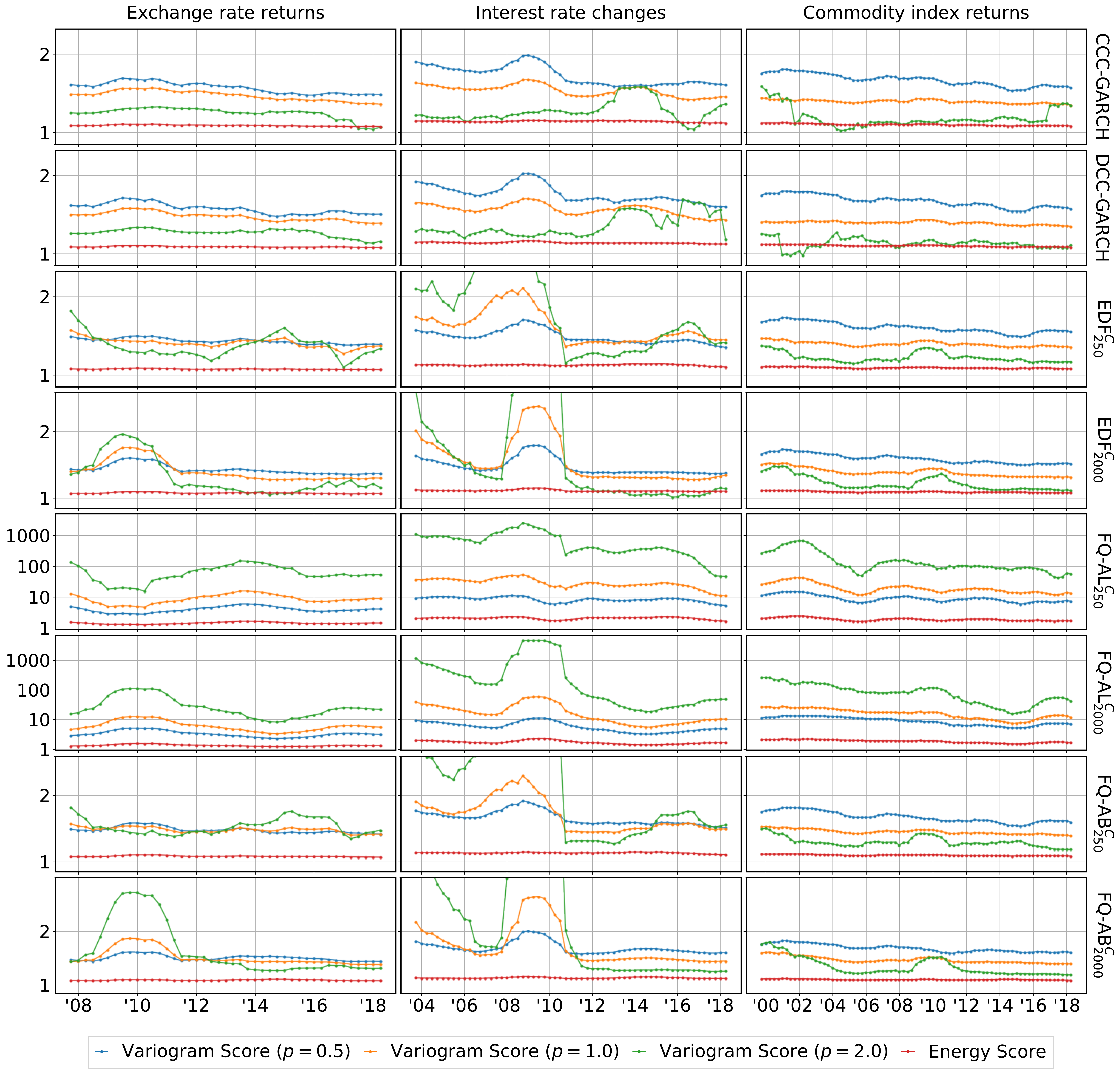} 
\end{centering}
{\footnotesize{}We display the discrimination heuristic of equation
(\ref{Equation:DisciminationHeuristic}) for all three data sets and
eight DGPs with a logarithmic scale. Scoring rules which separate
the scores of misspecified models and the DGP by a larger relative
distance are assigned higher values for the discrimination heuristic.
We smooth the discrimination heuristic with a moving average of 8
observations to improve the interpretability of the figure, but the
same patterns are present in case no smoothing is applied.}{\footnotesize\par}
\end{figure}

Through the consideration of multiple models, we go beyond ceteris
paribus sensitivities to obtain more general results. Our misspecified
models combine various misspecifications at once and are therefore
more similar to the settings under which the proper scoring rules
are applied in practice. Our generalized discrimination heuristic defined in \eqref{Equation:DisciminationHeuristic} allows us to test discrimination ability against multiple competing models at once. Note that we do not subtract the scores of the DGP
from those of the misspecified models in the numerator, but this does
not affect the rankings of the scoring rules. Figure~\ref{Figure:Discrimination} depicts our discrimination heuristic for each scoring rule over time with a logarithmic scale. Unlike Figure \ref{Figure:MeanScores}, results are illustrated for all DGPs (eight rows of plots) and all data sets (three columns of plots).

Figure \ref{Figure:Discrimination} shows a clear distinction between scoring rules in most cases, but the preferences
of the discrimination heuristic for the four rules can vary slightly depending on the data
set, the time period and the DGP chosen. Overall, there are several key features which emerge:
\begin{enumerate}
\item The energy score is always the scoring rule with the lowest discrimination
heuristic. This, again, is in accord with prior simulation studies
by \citet{pinson2013} and \citet{scheuerer2015}. Across all data
sets and DGPs, the energy score only receives an average discrimination
heuristic of $1.23$, compared to $2.79$, $5.30$ and $78.13$ for
the variogram score with $p=0.5$, $p=1$ and $p=2$. 
\item In all cases, the variogram score with $p=1$ is the scoring rule
with the second highest discrimination heuristic. 
\item The variogram score with $p=2$ achieves in some settings extremely
high values for the discrimination heuristic, but is also the only
scoring rule which receives values below 1. This occurs in commodity
index returns with DCC-GARCH as DGP. For those $t$, the model ranking
of the variogram score with $p=2$ is erroneous and multiple misspecified
models receive lower scores than the DGP; 
\item The scoring rule with the highest discrimination heuristic varies
depending on the choice of data and DGP but exhibits a pattern. In
most cases, the variogram score with $p=0.5$ has the highest discrimination
heuristic, but it is surpassed by the variogram score with $p=2$
during some periods and always when FQ-AL models are used as DGP. 
\end{enumerate}
The high discrimination heuristic of some variogram scores with $p=2$,
despite the poor performance in Figure \ref{Figure:MeanScores} can
be explained by Figure \ref{Figure:Variogram} and our discussion
on the effect of different choices of $p$ for the variogram score.
Generally, the variogram score with $p=2$ outputs a large range of
scores, some of which may be vastly larger in magnitude than others.
These outliers shift the sample mean in Figure \ref{Figure:MeanScores}
to a larger value than the sample 0.75-quantile and also affect the
discrimination heuristic to a similar extent. For instance, in exchange
rate returns with %
\mbox{%
DCC-GARCH%
} as DGP, the largest summand of equation (\ref{Equation:DisciminationHeuristic})
takes a value around 4,700. In comparison, the largest summand of
the energy score, variogram score with $p=0.5$ and $p=1$ are $17$,
$76$ and $141$ respectively.

The quadratic nature  of the variogram score with $p=2$
further amplifies large distances between models. Therefore, the variogram
score with $p=2$ achieves a particularly high discrimination heuristic
when the models are easily distinguishable. The cases where the variogram
score with $p=2$ have the highest discrimination heuristic mostly
correspond to two scenarios: 
\begin{enumerate}
\item Around the financial crisis in 2008, the differences of the distribution
forecasts become easier to distinguish. This is because models with
a calibration window of 2,000 observations are much less affected
by the abnormal values during the crisis in contrast to models with
a calibration window of only 250 observations. Hence, the distribution
forecasts may deviate more strongly between the competing models and
scoring rules may assign larger relative distances between the scores
of misspecified models and those of the DGP. 
\item Similarly, the use of FQ-AL as DGP also increases the relative distances
between the scores of the models. The Factor Quantile model typically produces
a much sharper forecast than that of alternative models and is therefore
more easily identifiable as DGP. 
\end{enumerate}
In those two cases, all scoring rules manage to quite clearly identify the
DGP from misspecified models, so the even larger relative distance
between the scores of the variogram score with $p=2$ arguably has no additional
benefit. Simultaneously, the scoring rule suffers from erroneous rankings,
despite having high discrimination heuristics in some settings. These
issues reveal clearly that the discrimination heuristic should be considered
as one possible indicator for the goodness of scoring rules, but by itself is
inadequate to quantify their discrimination ability. A large heuristic
of a scoring rule may not imply more robust or less erroneous rankings.
Therefore, a high discrimination heuristic between the models is not
useful unless it is accompanied by a low error rate, i.e. the percentage
of times choosing a misspecified model over the DGP.

\section{Conclusions}\label{sec:Conclusions}

In this paper we have conducted an extensive investigation of the discrimination ability of multivariate proper scoring rules, comparing the performance of four popular choices across a variety of time periods, financial data sets and competing stochastic models.  By rotating our DGP across a selection of eight realistic models with different static or dynamic joint distributions and different calibration windows, our simulation study provides a much less restrictive and more comprehensive set of results than the existing literature this area.  While results of course vary by scenario, overall trends and insights do emerge.  We observe that the energy score generally has poorer discrimination ability than the other scores, while the variogram score with $p=2$ can appear to provide strong discrimination in certain cases but on the other hand is prone to high error rates (failure to identify the correct DGP).  The variogram score with $p=0.5$ consistently maintains the lowest error rate, but is sometimes outperformed by the variogram score with $p=1$ in terms of discrimination ability since differences between models are magnified less.  On the whole, variogram score with $p=0.5$ seems to perform best in most settings, but another important conclusion is that one should typically consider multiple performance metrics and be aware of how different scoring rules tend to behave in different model settings. Such insights can undoubtedly be of great value when deciding which scoring rule to use in different multivariate forecasting applications. 

\cleardoublepage{\small{}\bibliographystyle{chicago}
\bibliography{BibTex}

\begin{thebibliography}{}

\bibitem[\protect\citeauthoryear{Alexander and Han}{Alexander and
  Han}{2020}]{Alexander2020}
Alexander, C. and Y.~Han (2020).
\newblock Static and dynamic models for multivariate distribution forecasts:
  Proper scoring rule tests of factor-quantile vs. multivariate garch models.
\newblock {\em ArXiV\/}~{\em 2004.14108}.

\bibitem[\protect\citeauthoryear{Amisano and Giacomini}{Amisano and
  Giacomini}{2007}]{amisano2007}
Amisano, G. and R.~Giacomini (2007).
\newblock Comparing density forecasts via weighted likelihood ratio tests.
\newblock {\em Journal of Business \& Economic Statistics\/}~{\em 25\/}(2),
  177--190.

\bibitem[\protect\citeauthoryear{{Bank of International Settlements}}{{Bank of
  International Settlements}}{2016}]{triennial2016}
{Bank of International Settlements} (2016, April).
\newblock Triennial {C}entral {B}ank survey: {F}oreign exchange turnover in
  {A}pril 2016.
\newblock Technical report.

\bibitem[\protect\citeauthoryear{Bao, Lee, and Salto{\u{g}}lu}{Bao
  et~al.}{2007}]{bao2007}
Bao, Y., T.-H. Lee, and B.~Salto{\u{g}}lu (2007).
\newblock Comparing density forecast models.
\newblock {\em Journal of Forecasting\/}~{\em 26\/}(3), 203--225.

\bibitem[\protect\citeauthoryear{Bauwens and Laurent}{Bauwens and
  Laurent}{2005}]{bauwens2005}
Bauwens, L. and S.~Laurent (2005).
\newblock A new class of multivariate skew densities, with application to
  generalized autoregressive conditional heteroscedasticity models.
\newblock {\em Journal of Business \& Economic Statistics\/}~{\em 23\/}(3),
  346--354.

\bibitem[\protect\citeauthoryear{Bickel}{Bickel}{2007}]{bickel2007}
Bickel, J.~E. (2007).
\newblock Some comparisons among quadratic, spherical, and logarithmic scoring
  rules.
\newblock {\em Decision Analysis\/}~{\em 4\/}(2), 49--65.

\bibitem[\protect\citeauthoryear{Bloomberg}{Bloomberg}{2017}]{bloomberg2017}
Bloomberg (2017, May).
\newblock The {B}loomberg commodity index family: {I}ndex methodology.
\newblock Technical report.

\bibitem[\protect\citeauthoryear{Bollerslev}{Bollerslev}{1986}]{bollerslev1986}
Bollerslev, T. (1986).
\newblock Generalized autoregressive conditional heteroskedasticity.
\newblock {\em Journal of Econometrics\/}~{\em 31\/}(3), 307--327.

\bibitem[\protect\citeauthoryear{Bollerslev}{Bollerslev}{1990}]{bollerslev1990}
Bollerslev, T. (1990).
\newblock Modelling the coherence in short-run nominal exchange rates: {A}
  multivariate generalized {ARCH} model.
\newblock {\em Review of Economics and Statistics\/}~{\em 72\/}(3), 498--505.

\bibitem[\protect\citeauthoryear{Breiman}{Breiman}{1996}]{breiman1996}
Breiman, L. (1996).
\newblock Bagging predictors.
\newblock {\em Machine Learning\/}~{\em 24\/}(2), 123--140.

\bibitem[\protect\citeauthoryear{Buja, Stuetzle, and Shen}{Buja
  et~al.}{2005}]{buja2005}
Buja, A., W.~Stuetzle, and Y.~Shen (2005).
\newblock Loss functions for binary class probability estimation and
  classification: {S}tructure and applications.

\bibitem[\protect\citeauthoryear{Cajigas and Urga}{Cajigas and
  Urga}{2006}]{cajigas2006}
Cajigas, J.-P. and G.~Urga (2006).
\newblock Dynamic conditional correlation models with asymmetric multivariate
  {L}aplace innovations.

\bibitem[\protect\citeauthoryear{Danielsson, James, Valenzuela, and
  Zer}{Danielsson et~al.}{2016}]{Danielsson2016}
Danielsson, J., K.~James, M.~Valenzuela, and I.~Zer (2016).
\newblock Model risk of risk models.
\newblock {\em Journal of Financial Stability\/}~{\em 23}, 79--91.

\bibitem[\protect\citeauthoryear{Dawid and Sebastiani}{Dawid and
  Sebastiani}{1999}]{dawid1999}
Dawid, P.~A. and P.~Sebastiani (1999).
\newblock Coherent dispersion criteria for optimal experimental design.
\newblock {\em Annals of Statistics\/}, 65--81.

\bibitem[\protect\citeauthoryear{Diebold, Gunther, and S}{Diebold
  et~al.}{1998}]{diebold1998}
Diebold, F.~X., T.~A. Gunther, and T.~A. S (1998).
\newblock Evaluating density forecasts, with applications to financial risk
  management.
\newblock {\em International Economic Review\/}~{\em 39}, 863--883.

\bibitem[\protect\citeauthoryear{Diebold and Mariano}{Diebold and
  Mariano}{1995}]{diebold1995}
Diebold, F.~X. and R.~S. Mariano (1995).
\newblock Comparing predictive accuracy.
\newblock {\em Journal of Business \& Economic Statistics\/}~{\em 13\/}(3),
  253--263.

\bibitem[\protect\citeauthoryear{Diks and Fang}{Diks and
  Fang}{2020}]{Diks2020531}
Diks, C. and H.~Fang (2020).
\newblock Comparing density forecasts in a risk management context.
\newblock {\em International Journal of Forecasting\/}~{\em 36\/}(2), 531--551.

\bibitem[\protect\citeauthoryear{Diks, Panchenko, Sokolinskiy, and van
  Dijk}{Diks et~al.}{2014}]{diks2014}
Diks, C., V.~Panchenko, O.~Sokolinskiy, and D.~van Dijk (2014).
\newblock Comparing the accuracy of multivariate density forecasts in selected
  regions of the copula support.
\newblock {\em Journal of Economic Dynamics and Control\/}~{\em 48}, 79--94.

\bibitem[\protect\citeauthoryear{Diks, Panchenko, and Van~Dijk}{Diks
  et~al.}{2011}]{diks2011likelihood}
Diks, C., V.~Panchenko, and D.~Van~Dijk (2011).
\newblock Likelihood-based scoring rules for comparing density forecasts in
  tails.
\newblock {\em Journal of Econometrics\/}~{\em 163\/}(2), 215--230.

\bibitem[\protect\citeauthoryear{Engle}{Engle}{1982}]{engle1982}
Engle, R.~F. (1982).
\newblock Autoregressive conditional heteroscedasticity with estimates of the
  variance of {U}nited {K}ingdom inflation.
\newblock {\em Econometrica: Journal of the Econometric Society\/}, 987--1007.

\bibitem[\protect\citeauthoryear{Engle}{Engle}{2001}]{engle2001}
Engle, R.~F. (2001).
\newblock {GARCH} 101: The use of {ARCH}/{GARCH} models in applied
  econometrics.
\newblock {\em Journal of Economic Perspectives\/}~{\em 15\/}(4), 157--168.

\bibitem[\protect\citeauthoryear{Engle}{Engle}{2002}]{engle2002}
Engle, R.~F. (2002).
\newblock Dynamic conditional correlation: {A} simple class of multivariate
  generalized autoregressive conditional heteroskedasticity models.
\newblock {\em Journal of Business \& Economic Statistics\/}~{\em 20\/}(3),
  339--350.

\bibitem[\protect\citeauthoryear{Feldmann, Scheuerer, and
  Thorarinsdottir}{Feldmann et~al.}{2015}]{feldmann2015}
Feldmann, K., M.~Scheuerer, and T.~L. Thorarinsdottir (2015).
\newblock Spatial postprocessing of ensemble forecasts for temperature using
  nonhomogeneous {Gaussian} regression.
\newblock {\em Monthly Weather Review\/}~{\em 143\/}(3), 955--971.

\bibitem[\protect\citeauthoryear{Gneiting, Balabdaoui, and Raftery}{Gneiting
  et~al.}{2007}]{gneiting2007a}
Gneiting, T., F.~Balabdaoui, and A.~E. Raftery (2007).
\newblock Probabilistic forecasts, calibration and sharpness.
\newblock {\em Journal of the Royal Statistical Society: Series B (Statistical
  Methodology)\/}~{\em 69\/}(2), 243--268.

\bibitem[\protect\citeauthoryear{Gneiting and Raftery}{Gneiting and
  Raftery}{2007}]{gneiting2007b}
Gneiting, T. and A.~E. Raftery (2007).
\newblock Strictly proper scoring rules, prediction, and estimation.
\newblock {\em Journal of the American Statistical Association\/}~{\em
  102\/}(477), 359--378.

\bibitem[\protect\citeauthoryear{Gneiting and Ranjan}{Gneiting and
  Ranjan}{2011}]{gneiting2011}
Gneiting, T. and R.~Ranjan (2011).
\newblock Comparing density forecasts using threshold-and quantile-weighted
  scoring rules.
\newblock {\em Journal of Business \& Economic Statistics\/}~{\em 29\/}(3),
  411--422.

\bibitem[\protect\citeauthoryear{Granger and Pesaran}{Granger and
  Pesaran}{2000}]{granger2000}
Granger, C. W.~J. and H.~M. Pesaran (2000).
\newblock A decision theoretic approach to forecast evaluation.
\newblock In {\em Statistics and Finance: An interface}, pp.\  261--278. World
  Scientific.

\bibitem[\protect\citeauthoryear{Hamill}{Hamill}{2001}]{hamill2001}
Hamill, T.~M. (2001).
\newblock Interpretation of rank histograms for verifying ensemble forecasts.
\newblock {\em Monthly Weather Review\/}~{\em 129\/}(3), 550--560.

\bibitem[\protect\citeauthoryear{Hyv{\"a}rinen}{Hyv{\"a}rinen}{2005}]{hyvarinen2005estimation}
Hyv{\"a}rinen, A. (2005).
\newblock Estimation of non-normalized statistical models by score matching.
\newblock {\em Journal of Machine Learning Research\/}~{\em 6\/}(Apr),
  695--709.

\bibitem[\protect\citeauthoryear{Jensen}{Jensen}{1968}]{jensen1968}
Jensen, M.~C. (1968).
\newblock The performance of mutual funds in the period 1945--1964.
\newblock {\em The Journal of Finance\/}~{\em 23\/}(2), 389--416.

\bibitem[\protect\citeauthoryear{Johnstone, Jose, and Winkler}{Johnstone
  et~al.}{2011}]{johnstone2011}
Johnstone, D.~J., V.~R.~R. Jose, and R.~L. Winkler (2011).
\newblock Tailored scoring rules for probabilities.
\newblock {\em Decision Analysis\/}~{\em 8\/}(4), 256--268.

\bibitem[\protect\citeauthoryear{Jordan, Kr{\"u}ger, and Lerch}{Jordan
  et~al.}{2017}]{jordan2017}
Jordan, A., F.~Kr{\"u}ger, and S.~Lerch (2017).
\newblock Evaluating probabilistic forecasts with the {R} package
  {scoringRules}.

\bibitem[\protect\citeauthoryear{Laio and Tamea}{Laio and
  Tamea}{2007}]{laio2007}
Laio, F. and S.~Tamea (2007).
\newblock Verification tools for probabilistic forecasts of continuous
  hydrological variables.
\newblock {\em Hydrology and Earth System Sciences Discussions\/}~{\em
  11\/}(4), 1267--1277.

\bibitem[\protect\citeauthoryear{Machete}{Machete}{2013}]{machete2013}
Machete, R.~L. (2013).
\newblock Contrasting probabilistic scoring rules.
\newblock {\em Journal of Statistical Planning and Inference\/}~{\em
  143\/}(10), 1781--1790.

\bibitem[\protect\citeauthoryear{Mandelbrot}{Mandelbrot}{1963}]{mandelbrot1963}
Mandelbrot, B.~B. (1963).
\newblock The variation of certain speculative prices.
\newblock {\em Journal of Business\/}~{\em 36}, 394--419.

\bibitem[\protect\citeauthoryear{Matheson and Winkler}{Matheson and
  Winkler}{1976}]{matheson1976}
Matheson, J.~E. and R.~L. Winkler (1976).
\newblock Scoring rules for continuous probability distributions.
\newblock {\em Management Science\/}~{\em 22\/}(10), 1087--1096.

\bibitem[\protect\citeauthoryear{Merkle and Steyvers}{Merkle and
  Steyvers}{2013}]{merkle2013}
Merkle, E.~C. and M.~Steyvers (2013).
\newblock Choosing a strictly proper scoring rule.
\newblock {\em Decision Analysis\/}~{\em 10\/}(4), 292--304.

\bibitem[\protect\citeauthoryear{Parry, Dawid, Lauritzen, et~al.}{Parry
  et~al.}{2012}]{parry2012proper}
Parry, M., A.~P. Dawid, S.~Lauritzen, et~al. (2012).
\newblock Proper local scoring rules.
\newblock {\em The Annals of Statistics\/}~{\em 40\/}(1), 561--592.

\bibitem[\protect\citeauthoryear{Pelagatti}{Pelagatti}{2004}]{pelagatti2004}
Pelagatti, M.~M. (2004).
\newblock Dynamic conditional correlation with elliptical distributions.

\bibitem[\protect\citeauthoryear{P\'erignon and Smith}{P\'erignon and
  Smith}{2010}]{perignon2010}
P\'erignon, C. and D.~Smith (2010).
\newblock The level and quality of value-at-risk disclosure by commercial
  banks.
\newblock {\em Journal of Banking and Finance\/}~{\em 34\/}(2), 362--377.

\bibitem[\protect\citeauthoryear{Pinson and Girard}{Pinson and
  Girard}{2012}]{pinson2012}
Pinson, P. and R.~Girard (2012).
\newblock Evaluating the quality of scenarios of short-term wind power
  generation.
\newblock {\em Applied Energy\/}~{\em 96}, 12--20.

\bibitem[\protect\citeauthoryear{Pinson and Tastu}{Pinson and
  Tastu}{2013}]{pinson2013}
Pinson, P. and J.~Tastu (2013).
\newblock Discrimination ability of the energy score.
\newblock Technical report.

\bibitem[\protect\citeauthoryear{Scheuerer and Hamill}{Scheuerer and
  Hamill}{2015}]{scheuerer2015}
Scheuerer, M. and T.~M. Hamill (2015).
\newblock Variogram-based proper scoring rules for probabilistic forecasts of
  multivariate quantities.
\newblock {\em Monthly Weather Review\/}~{\em 143\/}(4), 1321--1334.

\bibitem[\protect\citeauthoryear{Sta{\"e}l~von Holstein}{Sta{\"e}l~von
  Holstein}{1970}]{vonHolstein1970}
Sta{\"e}l~von Holstein, C.-A.~S. (1970).
\newblock Measurement of subjective probability.
\newblock {\em Acta Psychologica\/}~{\em 34}, 146--159.

\bibitem[\protect\citeauthoryear{Sz{\'e}kely}{Sz{\'e}kely}{2003}]{szekely2003}
Sz{\'e}kely, G.~J. (2003).
\newblock E-statistics: {T}he energy of statistical samples.
\newblock {\em Bowling Green State University, Department of Mathematics and
  Statistics Technical Report\/}~{\em 3\/}(05), 1--18.

\bibitem[\protect\citeauthoryear{Tsui and Yu}{Tsui and Yu}{1999}]{tsui1999}
Tsui, A.~K. and Q.~Yu (1999).
\newblock Constant conditional correlation in a bivariate {GARCH} model:
  {E}vidence from the stock markets of {C}hina.
\newblock {\em Mathematics and Computers in Simulation\/}~{\em 48\/}(4--6),
  503--509.

\bibitem[\protect\citeauthoryear{Winkler}{Winkler}{1971}]{winkler1971}
Winkler, R.~L. (1971).
\newblock Probabilistic prediction: {S}ome experimental results.
\newblock {\em Journal of the American Statistical Association\/}~{\em
  66\/}(336), 675--685.

\bibitem[\protect\citeauthoryear{Winkler}{Winkler}{1977}]{winkler1977}
Winkler, R.~L. (1977).
\newblock Rewarding expertise in probability assessment.
\newblock In {\em Decision Making and Change in Human Affairs}, pp.\  127--140.
  Springer.

\bibitem[\protect\citeauthoryear{Winkler}{Winkler}{1996}]{winkler1996}
Winkler, R.~L. (1996).
\newblock Scoring rules and the evaluation of probabilities.
\newblock {\em Test\/}~{\em 5\/}(1), 1--60.

\bibitem[\protect\citeauthoryear{Ziel and Berk}{Ziel and
  Berk}{2019}]{ziel2019multivariate}
Ziel, F. and K.~Berk (2019).
\newblock Multivariate forecasting evaluation: On sensitive and strictly proper
  scoring rules.
\newblock {\em arXiv preprint arXiv:1910.07325\/}.

\end{thebibliography}
}{\small\par}
\end{document}